\documentclass[journal]{IEEEtran}
\usepackage{afterpage}
\usepackage{bm}
\usepackage{comment}

\usepackage{cite}
\usepackage{color}

\ifCLASSINFOpdf
   \usepackage[pdftex]{graphicx}
   \graphicspath{{../pdf/}{../jpeg/}}
   \DeclareGraphicsExtensions{.pdf,.jpeg,.png}
\else
   \usepackage[dvips]{graphicx}
   \DeclareGraphicsExtensions{.pdf}
\fi

\usepackage{algorithm}
\usepackage{algpseudocode}

\usepackage{amsmath,amssymb,here,boxedminipage}
\usepackage{amsthm}
\usepackage{array}

\ifCLASSOPTIONcompsoc
  \usepackage[caption=false,font=normalsize,labelfont=sf,textfont=sf]{subfig}
\else
  \usepackage[caption=false,font=footnotesize]{subfig}
\fi

\usepackage{fixltx2e}

\usepackage{url}
\begin{document}

\title{Detailed Dynamic Model of Antagonistic PAM System and its Experimental Validation:\\ Sensor-less Angle and Torque Control with UKF}
\author{Takaya Shin,~\IEEEmembership{Student Member,~IEEE},~Takumi Ibayashi,~\IEEEmembership{Student Member,~IEEE},~\\and Kiminao Kogiso,~\IEEEmembership{Member,~IEEE}
\thanks{T. Shin and K. Kogiso are with the Department of Mechanical and Intelligent Systems Engineering, 
The University of Electro-Communications,
1-5-1 Chofugaoka, Chofu, Tokyo 1828585, Japan.
e-mail: \{shintakaya, kogiso\}@uec.ac.jp.}%
\thanks{T. Ibayashi is with Safie, Inc.}}

\maketitle
\begin{abstract}
This paper proposes a detailed nonlinear mathematical model of an antagonistic pneumatic artificial muscle (PAM) actuator system for estimating the joint angle and torque using an unscented Kalman filter (UKF).
The proposed model is described in a hybrid state-space representation. It includes the contraction force of the PAM, joint dynamics, fluid dynamics of compressed air, mass flows of a valve, and friction models.
A part of the friction models is modified to obtain a novel form of Coulomb friction depending on the inner pressure of the PAM.
For model validation, offline and online UKF estimations and sensor-less tracking control of the joint angle and torque are conducted to evaluate the estimation accuracy and tracking control performance.
The estimation error is less than 7.91 \%, and the steady-state tracking control performance is more than 94.75 \%.
These results confirm that the proposed model is detailed and could be used as the state estimator of an antagonistic PAM system.
\end{abstract}


\IEEEpeerreviewmaketitle

\section{Introduction}
\IEEEPARstart{T}{he} McKibben pneumatic artificial muscle (PAM) actuator system has a high strength-to-weight ratio and excellent flexibility.
It consists of an internal rubber tube surrounded by a cylindrical mesh braided by inextensible threads. 
Both ends of the two-layered tube are closed by caps to retain the cylindrical form, and one cap has a connector to supply compressed air. 
The tube diameter increases when compressed air is supplied; further, the long axis shortens because of the inextensible threads. In this manner, the PAM generates a contraction force, and 
it returns to its original shape through its elasticity when the compressed air is released from the inner tube.
Because a PAM can generate only a contraction force, 
an antagonistic structure consisting of two PAMs in parallel, with one connected on each side via a rotational joint, is often used.
A PAM is a suitable actuator for devices such as assist robots, nursing care robots, rehabilitation orthoses, and other robots that are often in contact with humans.
However, a PAM has high nonlinearity owing to the pressure dynamics and friction, and therefore, the modeling and control of an antagonistically structured PAM actuator system are difficult and challenging.
Several studies have conducted numerical simulations of PAM behaviors, such as the geometric-model-based approach \cite{2011_Borzikova} and Hill-model-based approach \cite{MHM1,MHM2}. Studies have also investigated PAM control methods such as a sliding mode control \cite{2018_Cao}, adaptive control \cite{2020_Li}, and guaranteed-cost control \cite{2013_Amato} by considering the modeling uncertainty and high nonlinearity of PAM actuators.

The characteristic compliance of a PAM actuator system plays an important role in its flexibility\cite{2011_Choi,2015_Saito}.
\cite{2006_Tsuji} noted that the control bandwidth becomes smaller at low pressure, and therefore, PAM control becomes difficult.
A nonlinear dynamic model of the PAM actuator system can overcome the difficulty of control in the lower-pressure range and enable maintaining the high compliance of the actuator system.
For example, a sliding mode controller with a pressure model was used to achieve angle-compliance control of an antagonistic PAM actuator and was applied to rehabilitation orthoses\cite{2018_Cao}; however, this study did not consider the hysteresis.
Hysteresis increases the complexity of the actuator system, and when it is left unmodeled, it makes the control of PAMs difficult\cite{06_Davis}.
Indeed, hysteresis compensation has been shown to improve the control performance\cite{2012_Schreiber, 2019_Song}.
Furthermore, the contraction force plays a crucial role in the PAM actuator system, and a contraction force model is associated with PAM pressure through a static and nonlinear function\cite{2018_Survey}.
However, it is difficult to identify the mechanism by which a contraction force is generated because of the large number of parameters and the complex structure of a PAM\cite{2012_Sui}. Therefore, 
some studies have employed an empirical model (experimentally approximated function) \cite{2005_Hildebrandt, 2012_expmodel, 2012_polymodel, 2015_polymodel}.
Overall, a model-based approach is effective for developing and improving a PAM actuator system.

With a precise model of the PAM actuator system, the measured pressure information can be used to estimate the joint angle and torque behaviors of this system.
The use of a force sensor (i.e., a load cell) and an encoder enabled torque and stiffness control of a PAM joint actuator\cite{2010_stiffness, 2017_Zhao, 2019_Ugurl}.
Among sensor-less approaches, one study \cite{2015_polymodel} showed that the use of a force map instead of a force sensor helps to estimate the joint torque of the PAM system, and another study applied a force-sensor-less approach to achieve torque control of an antagonist PAM actuator system under a fixed joint angle\cite{sensorless_PAM}.
However, conventional studies still use a sensor, such as an encoder, for measuring the joint angle. 
Simply estimating the joint angle could further reduce the time required for designing and constructing a PAM actuator system.
An unscented Kalman filter (UKF) \cite{ukf_2000} is used to estimate the state of nonlinear systems from control inputs and measured data.
It was experimentally demonstrated in \cite{2017_Kodama} that the UKF enables the estimation of the contraction ratio for a practical single PAM system. 
Further, it was shown in \cite{2018_Yokoyama}  that the UKF helps achieve position-sensor-less control for a PAM system.
Therefore, a sensor-less control scheme shows promise for developing a practical antagonistic PAM actuator system with low weight and high flexibility.

In this light, this paper proposes a detailed nonlinear model of an antagonistic PAM system that is actuated by a proportional directional control valve (PDCV).
The proposed model has two inputs---control commands to the two PDCVs---and four outputs---joint angle, torque, and two PAM pressures. 
This model is constructed by using the geometric relations and motion equations of two PAMs and a rotational joint with friction, fluid dynamics, a contraction force, and compressed air flows in a PDCV in the same manner as the single PAM modeled in previous studies\cite{2011_Itto}\cite{2015_Urabe}.
The proposed model employs the Kikuuwe model based on an implicit Euler integration for friction\cite{2006kiku} and the Itto model that associates an input voltage to a PDCV with its open ratio, a model parameter\cite{2011_Itto}.
A remarkable feature of the proposed model is that it captures systems behaviors in an absolute pressure range of 200--700 kPa, which enables adjusting the compliance sufficiently.
In addition, our study provides a classification result for model parameters that indicates how they can be estimated using measured data.
Furthermore, this study evaluates the proposed model by confirming the angle and torque control performance of an antagonistic PAM actuator system with an UKF-based estimator. 
This UKF uses the proposed model to estimate the angle and torque from only the measured pressure through the following procedure.
First, the angle and torque estimation results are compared with/without the UKF in a numerical simulation that uses experimental test data measured in advance.
Second, the UKF-based estimator is implemented in the antagonistic PAM actuator system and an online estimation of the angle and torque is conducted to evaluate estimation errors.
Finally, a UKF-based sensor-less proportional-integral (PI) feedback control system is constructed to evaluate whether the proposed model provides good control performance.
The experimental validation reveals that the proposed nonlinear model of the antagonistic PAM actuator system is valid and detailed enough to be applied to model-based sensor-less control. 
This study makes the following main contributions: it is the first study to implement a UKF into an antagonistic PAM actuator system to achieve joint angle and torque estimations, and it provide novel forms of the pressure-dependent Coulomb friction and the frictional force acting on the shaft for facilitating the improvement of the UKF's state estimation.

The reminder of this paper is organized as follows:
Section \ref{sec:model} introduces the antagonistic PAM system and its mathematical model.
Section \ref{sec:ModelValidation} describes the validation of the proposed model using UKF in both offline and online estimations.
Section \ref{sec:CS} presents the experimental results of the UKF-based angle and torque control; these results confirm that the joint angle and torque can be used for feedback control instead of sensors.
Finally, Section \ref{sec:Conclusion} presents the conclusions of this study.

\section{Antagonistic PAM System and its Mathematical Model}\label{sec:model}

\subsection{Experimental Setup of Antagonistic PAM System}
\begin{figure}[t]
	\centering
	\subfloat[Photograph of antagonistic PAM system.]{\includegraphics[width=0.75\hsize]{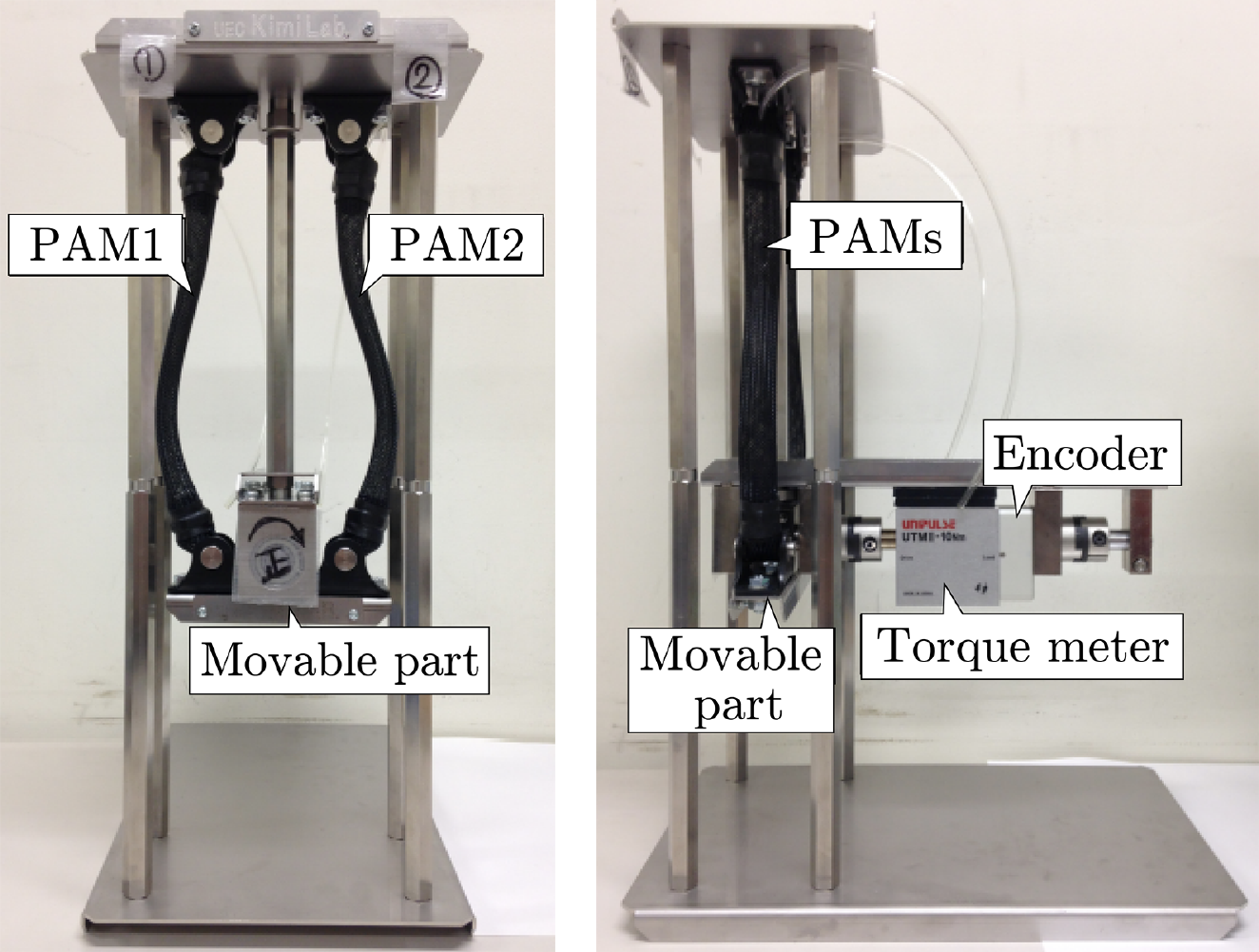}
	\label{subfig:atgsys}}\\
	\subfloat[Schematic of antagonistic PAM system.]{\includegraphics[width=0.75\hsize]{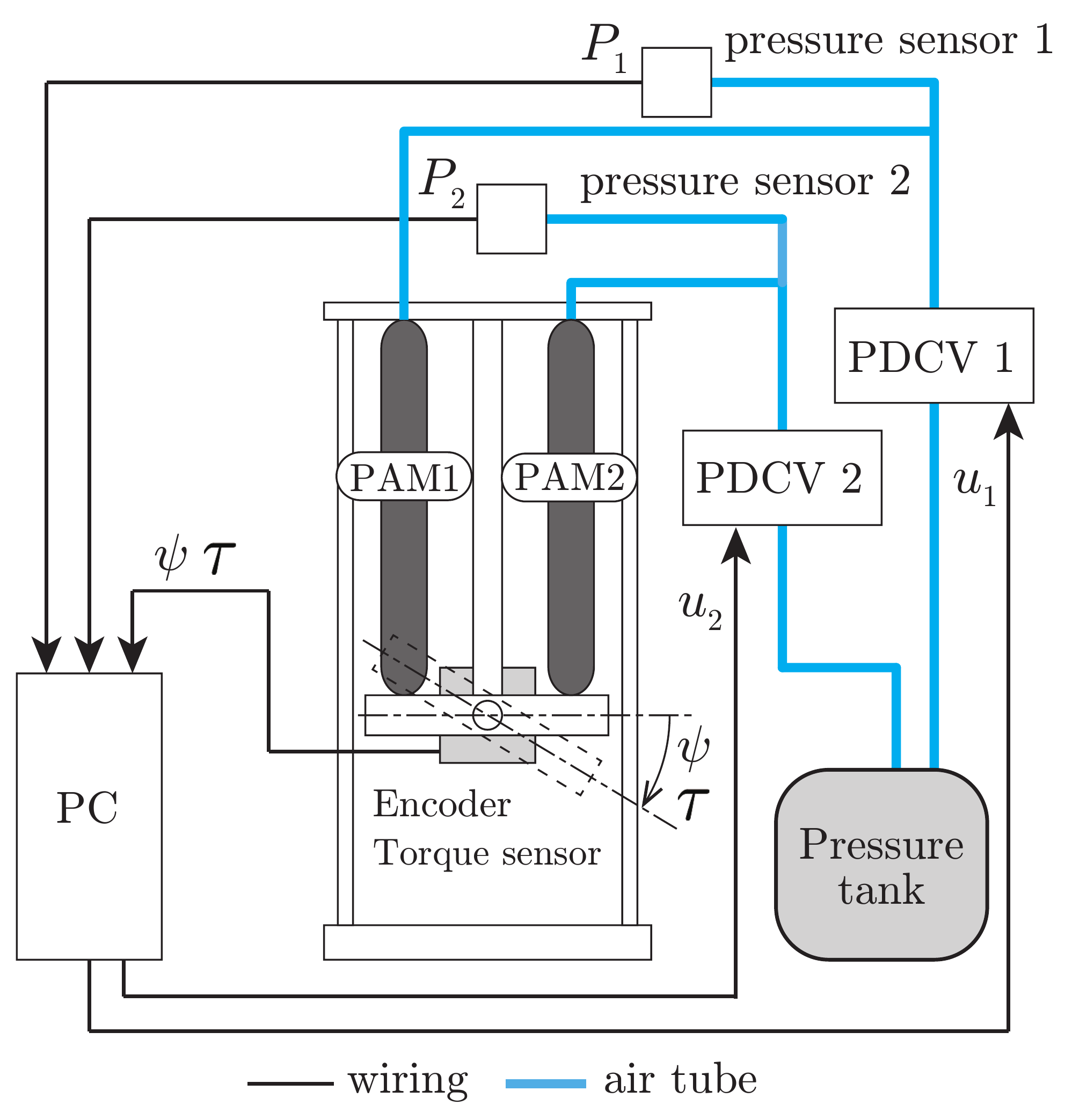}%
	\label{subfig:eqatg}}
	\caption{Experimental antagonistic PAM system.}
	\label{fig:setup}
\end{figure}

The antagonistic PAM system is a joint actuator powered by two PAMs.
Figs. \ref{fig:setup}\subref{subfig:atgsys} and \subref{subfig:eqatg} respectively show the appearance and the structure of the antagonistic PAM system.
This system consists of two PAMs~(Airmuscle, Kanda Tsushin Kogyo), two PDCVs~(5/3-way valve, FESTO), an air tank~(6-25, JUN-AIR), sensors, and a control PC.
One side of each PAM is connected with a movable part.
The tank stores compressed air and is connected to the PDCVs and PAMs by air tubes.
The pressure controlled by the PDCVs drives the PAMs to make the movable part rotate with a seesaw motion.
A rotary encoder and a torque meter (TM II-10 Nm(R), UNIPULSE) are used to measure the joint angle and the torque, respectively, and two pressure sensors~(E8F2-B10C, OMRON) are used to measure the inner pressure of PAMs.
The system inputs are the voltage signals to the two PDCVs ($u_1$ and $u_2$), and the measured values are the joint angle $\psi$, torque $\tau$, and inner pressures of the two PAMs ($ P_1$ and $P_2$).
The PC has a 3.2~GHz CPU and 8~GB RAM, and the operating system used is Ubuntu~12.04 with the preemption-patched Xenomai~2.6.2.1.
The sampling period of the PC was set to 1 ms. 
The range of the rotation angle is $\pm$25$^\circ$, and the range of the output torque is $\pm$3.0 Nm.

\subsection{Nonlinear Mathematical Model}\label{subsec:Model}
The state-space model of the PAM system in consideration of noise is expressed as follows:
	\begin{subequations}
		\begin{align}
			\dot{x}(t) &= f_\sigma(x(t), u(t)) + v(t) \hspace*{3ex}{\sf if} \ \ x(t) \in \mathcal{X}_{\sigma}, \label{eq:pam_a} \\
			y(t) &= h(x(t)) + w(t), \label{eq:pam_b}
		\end{align}
		\label{eq:pam}
	\end{subequations}
\!\!where $t \in \mathbb{R}_{\geq 0}$ is the time; 
$u:=[u_1 \ u_2]^T\in{\mathcal U}\subset \mathbb{R}^2$ is a control input with input voltages $u_1$ and $u_2$ to the PDCAs connected to the PAMs, respectively; and 
${\mathcal U}:=[0,\,10]^2$ is a set of allowable control inputs. 
The state variable is $x:=[\psi \ \dot{\psi} \ P_1 \ P_2 ]^\mathrm{T} \in \mathbb{R}^4$.
The output variable is $y:=[\psi \ P_1 \ P_2 \ \tau]^\mathrm{T} \in \mathbb{R}^4$.
$v$ and $w$ are the process noise and observation noise, respectively.
$f_{\sigma}:\mathbb{R}^4\rightarrow \mathbb{R}^4$ is a nonlinear function with 18 subsystems; it switches according to if-then rules.
${\mathcal X}_\sigma: =\{x \in \mathbb{R}^4 | \Psi_{\sigma}(x)>0 \}$ are the state sets, where $\sigma\in \Sigma:=\{1, 2, \cdots, 18\}$ is the index of the subsystem.
$\Psi_{\sigma}(x)$ is a function derived from the modes in the form of if-then rules.
The function $h:\mathbb{R}^4\rightarrow \mathbb{R}^4$ is an observation equation.
The model is obtained by summarizing the following components of the PAM system:

\subsubsection{Geometric equation}
The length of the two PAMs is denoted by $l_1$ and $l_2$ and is given by
	\begin{align}
		l_1(t)=L_0-\Delta L (t) ,\ l_2(t)=L_0+\Delta L (t),
		\label{eq:l}
	\end{align}
where $\Delta L (t)\approx r\sin \psi(t)$ is the vertical displacement of the PAM length; $r$, the radius of the seesaw; and $L_0$, the PAM length at the horizontal position of the seesaw. Indeed, rotation of the seesaw causes vertical and horizontal displacements; in this study, the horizontal displacement is ignored because it negligibly affects the PAM length.
As a model of the PAM volume, this study employs a quadratic polynomial function of the contraction rate, as is used in \cite{KIMURA19971385,MINH2010402}, because an accurate model of the PAM volume is too complex to obtain analytically. 
The PAM volume is given by
	\begin{eqnarray}
		V_i(t)= D_1 l_i(t) ^2 + D_2 l_i(t) + D_3, \quad \forall\,i\in{\mathcal I}:=\{1,2\},
		\label{eq:V}
	\end{eqnarray}
where $D_1$, $D_2$, and $D_3$ are experimentally determined coefficients.
The time derivate of \eqref{eq:V} is given as $\dot{V}_i(t)= (2D_1 l_i(t) + D_2) \dot{l}_i(t)$~($\forall\,i\in{\mathcal I}$),
where $\dot{l}_i$ can be calculated by differentiating \eqref{eq:l}.

\subsubsection{Fluid dynamics}\label{subsec:FD}
The change in the inner pressure of the PAM can be modeled thermodynamically.
The energy balance in a PAM is used to formulate the pressure change rate\cite{2011_Itto, Richer2001AHP}:
	\begin{eqnarray}
		\dot{P}_i(t) = k_1\frac{RT}{V_i(t)}m_i(t)-k_2\frac{\dot{V}_i(t)}{V_i(t)}P_i(t),
		\label{eq:P}
	\end{eqnarray}
where $m_i$ is the mass flow rate of compressed air streaming from the PDCV to a PAM, $k_1$ and $k_2$ are polytropic indexes, 
$R$ is the gas constant, and $T$ is the absolute temperature of air.
Hereafter, the subscript $i$ is omitted to simplify the notations.
The PDCV is characterized by using the mass flow rate that is expressed as\cite{2011_Itto, Richer2001AHP}
	\begin{eqnarray}
		 m(t) = \alpha(t) m_{\rm in}(t) - (1- \alpha(t) ) m_{\rm out}(t),
		\label{eq:m}
	\end{eqnarray}
where $m_{\rm in}$ and $m_{\rm out}$ are respectively the mass flow rates entering and leaving the intake port of the PDCV.
$\alpha\in[0,1]$ is equivalent to the open rate of the valve and depends on the voltage signal $u$.
$\alpha$ is written as a function of $u$ as $\alpha=\kappa(u)$, where $\kappa$ is a monotonically increasing function with respect to $u\in\mathcal{U}:=[\kappa^{-1}(0),\ \kappa^{-1}(1)]$.
$\alpha$ characterizes the inner pressure of the static PAM; the function is described in Section \ref{subsec:modelpara}.
$m_{\rm in}$ and $m_{\rm out}$ are respectively expressed as
	\begin{eqnarray*}
		{\footnotesize
		m_{\rm in}(t) = \left \{ 
		\begin{array}{l}
			A_0 \displaystyle\frac{P_{tank}}{\sqrt{T}}\sqrt{\frac{k}{R}\left( \frac{2}{k+1} \right)^{\frac{k+1}{k-1}}}\\[3ex]
			\hspace{8em} {\sf if} \hspace{1em} P(t) \leq P_{tank}\left( \frac{2}{k+1} \right)^{\frac{k}{k-1}}, \\[3ex]
			A_0 \displaystyle\frac{P_{tank}}{\sqrt{T}} \sqrt{\frac{2k}{R(k-1)}} \left( \frac{P(t)}{P_{tank}} \right)^{\frac{1}{k}} \sqrt{1-\left( \frac{P(t)}{P_{tank}} \right)^\frac{k-1}{k}}\\[3ex]
			\hspace{8em} {\sf if} \hspace{1em} P(t) >  P_{tank}\left( \frac{2}{k+1} \right)^{\frac{k}{k-1}},
		\end{array} \right.
		}
	\end{eqnarray*}
	\begin{eqnarray*}
		{\footnotesize
			m_{\rm out}(t) = \left \{ 
		\begin{array}{l}
			A_0 \displaystyle\frac{P(t)}{\sqrt{T}}\sqrt{\frac{k}{R}\left( \frac{2}{k+1} \right)^{\frac{k+1}{k-1}}}\\[3ex]
			\hspace{8em} {\sf if} \hspace{1em} P_{\rm out} \leq P(t)\left( \frac{2}{k+1} \right)^{\frac{k}{k-1}}, \\[3ex]
			A_0 \displaystyle\frac{P(t)}{\sqrt{T}} \sqrt{\frac{2k}{R(k-1)}} \left( \frac{P_{\rm out}}{P(t)} \right)^{\frac{1}{k}} \sqrt{1-\left( \frac{P_{\rm out}}{P(t)} \right)^\frac{k-1}{k}}\\[3ex]
			\hspace{8em} {\sf if} \hspace{1em} P_{\rm out} >  P(t)\left( \frac{2}{k+1} \right)^{\frac{k}{k-1}}.
		\end{array} \right.
		}
	\end{eqnarray*}
\normalsize
Mass flow loss is seen at the orifice, and it differs at the orifice for entering and leaving the PDCV.
This study considers a different orifice area $A_0$ according to the flow direction; specifically, $A_0=A_1$ if $m(t)>0$ and $A_0=A_2$ if $m(t)\leq0$.

\subsubsection{Contraction force}
\begin{figure}[t]
	\centering
	\subfloat[$P-F$ relationship of the PAM.]{\includegraphics[width=0.8\hsize]{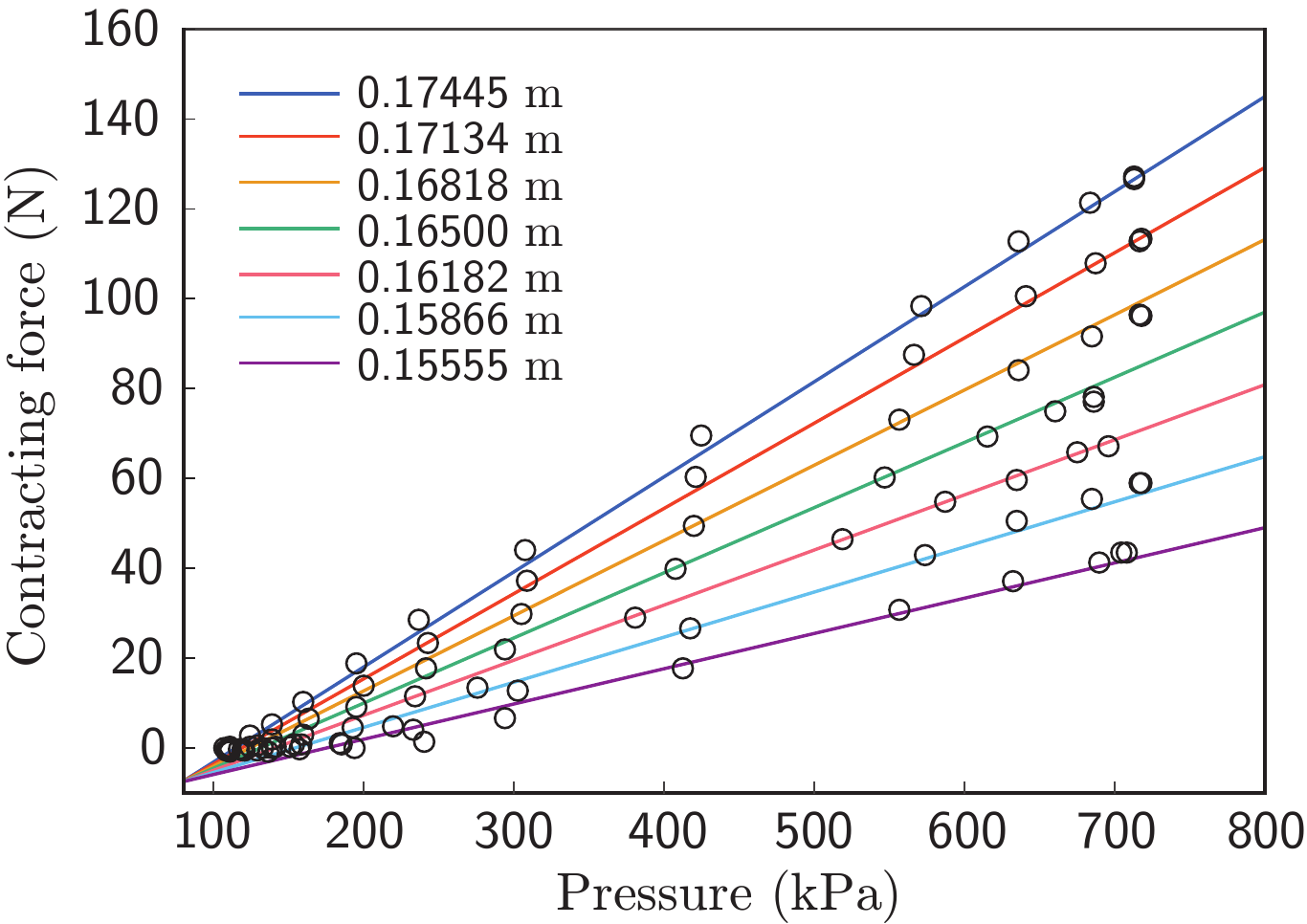}
	\label{subfig:PF_PAM1}}\\
	\subfloat[$v-l$ and $w-l$ relationships of the PAM.]{\includegraphics[width=0.8\hsize]{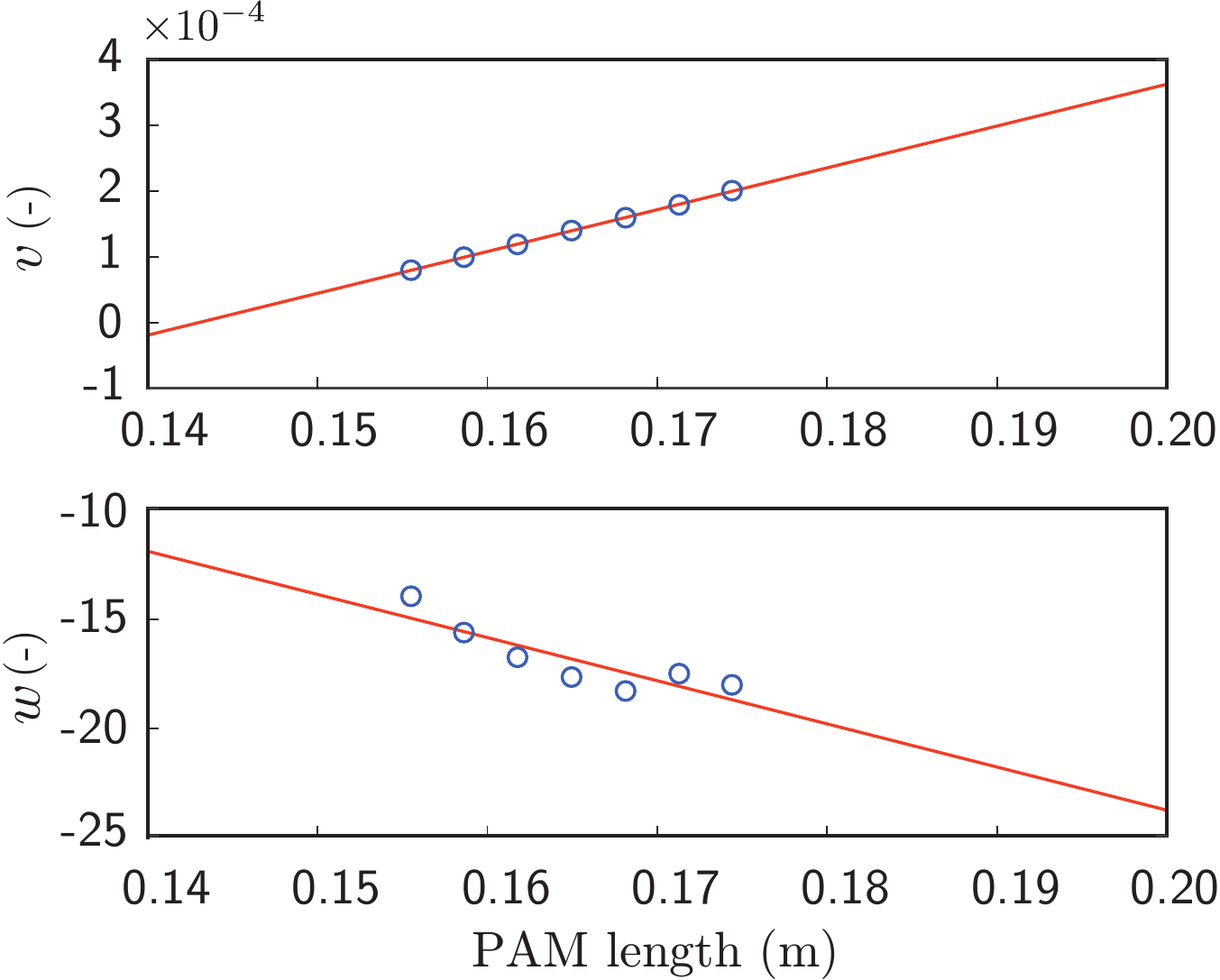}%
	\label{subfig:vw_PAM1}}
	\caption{Static characteristics with regard to $P$, $F$, and $l$ of the PAMs.}
	\label{fig:linear_model}
\end{figure}

Because a contraction force is statically associated with the inner pressure under a fixed PAM length\cite{2011_Minh},
this study experimentally clarifies the relationship between the inner pressure and the contracting force, as shown in Fig.\ \ref{fig:linear_model}\subref{subfig:PF_PAM1}. 
In this figure, the black circles indicate experimental data, and the relationship for each length is approximated as a linear function of pressure.
The linear function outputs a contraction force $F$ and is expressed as
	\begin{align}
		F(P(t))=v P(t)+w, 
		\label{eq:force}
	\end{align}
where $v$ and $w$ are respectively the slope and intercept of the line graph.
The fitted results obtained using \eqref{eq:force} are indicated by the colored solid lines in Fig.\ \ref{fig:linear_model}\subref{subfig:PF_PAM1}.
This figure shows that $v$ and $w$ depend on the PAM length, and therefore, they are described as a function of $l$.
For the relations between $l$ and $v$ and between $l$ and $w$, 
the blue circles in Fig.\ \ref{fig:linear_model}\subref{subfig:vw_PAM1} indicate the computed pairs with respect to $v$ and $w$; these are seen to be linear in length.
	\begin{subequations}
		\begin{align}
			v(l(t)) &= p_{v_1} l(t) + p_{v_2},\label{eq:pamfrc1}\\
			w(l(t)) &= p_{w_1} l(t) + p_{w_2}.\label{eq:pamfrc2}
		\end{align}
		\label{eq:pamfrc}
	\end{subequations}
\!\!The fitting results obtained using \eqref{eq:pamfrc1} and \eqref{eq:pamfrc2} are indicated by the solid red lines in Fig.\ \ref{fig:linear_model}\subref{subfig:vw_PAM1}.

\subsubsection{Joint dynamics}
\begin{figure}[tb]
	\centering
			\includegraphics[width=0.95\hsize]{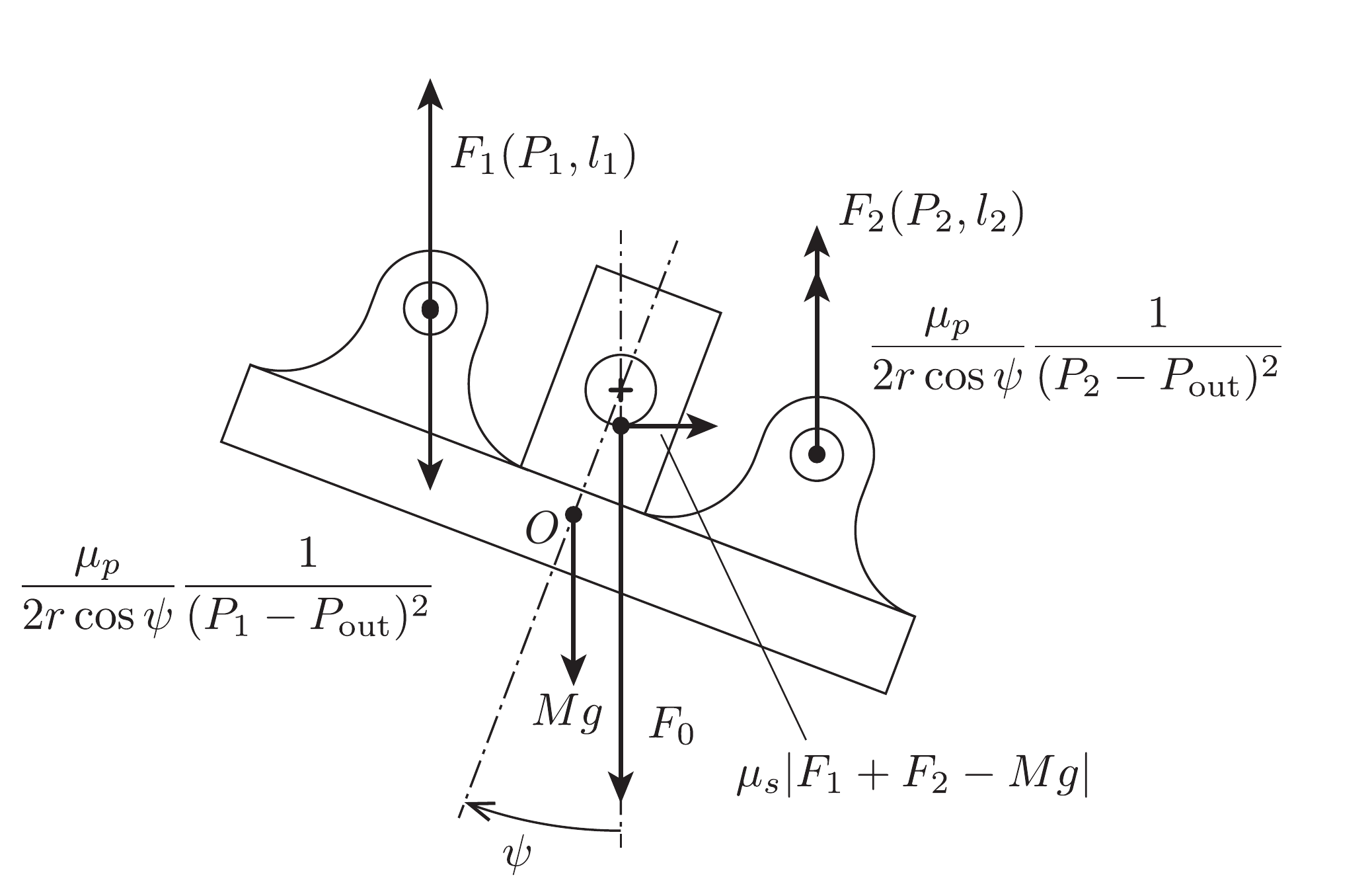}
			\caption{Illustration of forces acting on the joint.}
			\label{fig:ssfric}
	\end{figure}

The equation of the seesaw motion is written as
	\begin{align}
		J \ddot{\psi}(t) = \tau(t) - T_f(t) - k_s\psi(t),
		\label{eq:ssem2}
	\end{align}
where $J$ is the moment of inertia of the seesaw, $\tau$ is the torque generated by PAMs, $T_f$ is the resistance torque caused by friction, and $k_s$ is the experimentally determined torque coefficient caused by the seesaw motion with respect to the angle.
Considering the geometric relationship shown in Fig.\ \ref{fig:ssfric}, $\tau$ is expressed as follows:
	\begin{align}
		\tau(t) = r\cos \psi(t) \left( F_1(t) - F_2(t) \right),
		\label{eq:tauPAM}
	\end{align}
\!\!The resistance torque is given by a discrete-time friction model\cite{2006kiku}:
	\begin{eqnarray}
		T_f(k) = \begin{cases}
			\dfrac{(T_s(k) + T_p(k)){\rm sgn}(\dot{\psi}(k))+c_s \dot{\psi}(k)}{1+Zc_s}, & \\[1.5ex]
			\hspace*{9ex} {\sf if} \,\ \left| \dot{\psi}(k) \right| > Z(T_s(k)+T_p(k)), \\[2ex]
			\dfrac{\dot{\psi}(k)}{Z}\qquad {\sf if}\,\ \left| \dot{\psi}(k) \right| \leq Z(T_s(k)+T_p(k)),
		\end{cases}
		\label{eq:ff}
	\end{eqnarray}
\!\!where $\dot{\psi}(k)$ is the rotational velocity at step $k\in\mathbb{Z}_{\geq 0}$, where $t=kT_{\rm stp}$ holds with a sampling time $T_{\rm stp}$.
The velocity is calculated using the Euler method $\dot{\psi}(k)=\dot{\psi}(k-1) + ZT_0(k)$, where $Z=J/T_{\rm stp}$ and 
$T_0$ is the torque except for the frictional force at step $k$, that is, $T_0(k) = \tau(k) - k_s\psi(k)$.
$T_s$ and $T_p$ are the frictional forces acting on the shaft and rubber tube, respectively. 
From Coulomb's law of friction, $T_s$ is proportional to the force applied to the shaft as follows:
	\begin{eqnarray*}
		T_s(k) = r_p \mu_s |F_1(k) + F_2(k) - Mg|,
		\label{eq:ssf}
	\end{eqnarray*}
where $r_p$ is the shaft radius and $\mu_s$ is a constant coefficient.
$T_p$ denotes the frictional force effect on a tube and a mesh inside the PAM.
In \cite{2015_Urabe}, it was observed that if the PAM inner pressure increases, then $T_p$ decreases. Therefore, this study makes an observation in the following novel form:
	\begin{eqnarray*}
		T_p(k) = \mu_p\left( \frac{1}{(P_1(k)-P_{\rm out})^2}+\frac{1}{(P_2(k)-P_{\rm out})^2} \right),
	\end{eqnarray*}
where $\mu_p$ is a constant coefficient.

\subsubsection{State-space model of antagonistic PAM system}
The state-space equation $f_{\sigma}(x(k))$ \eqref{eq:pam_a} and the output equation $h(x(k))$ \eqref{eq:pam_b} are described as follows:
	\begin{subequations}
		\begin{align}
			f_{\sigma}(x(t),u(t))&=\left[\begin{matrix}
				\dot{\psi}(t)\\
				(\tau(t) - T_f(t) - k_s\psi(t))/J\\
				k_1\frac{RT}{V(t)}m_1(t) - k_2\frac{\dot{V}(t)}{V(t)}P_1(t)\\
				k_1\frac{RT}{V(t)}m_2(t) - k_2\frac{\dot{V}(t)}{V(t)}P_2(t)
				\end{matrix}\right],
			\label{eq:state_eq}\\
			\nonumber \\ 
			h(x(t))&=\left[\begin{matrix} \psi(t) \\ P_1(t) \\ P_2(t) \\ r\cos \psi(t)(F_1(t)-F_2(t)) \end{matrix}\right].
			\label{eq:kansoku}
		\end{align}
		\label{eq:pamfrc}
	\end{subequations}

\subsection{Model Parameters}\label{subsec:modelpara}
Table \ref{tbl:variables} lists the model parameters of the antagonistic PAM system; these include directly measurable and estimated parameters.
From the steady-state analysis of the model given by \eqref{eq:state_eq}, the following feature is stated.
\newtheorem{prop}{Proposition} 
\begin{prop}\label{prop1}
The model parameters (shown in Table \ref{tbl:variables}) that need to be estimated are divided into two groups: 
$A_0$, $k_1$, and $k_2$ affect transient responses, 
and $T_p'$ and $\mu_s$ affect steady-state responses.
\end{prop}
\begin{IEEEproof}
		Let the steady-state parameters of pressure $P_i(t)$ and $\psi(t)$ be $\bar{P_i}\in[P_{\rm out},P_{\rm in}]$ and $\bar{\psi}\in\mathbb{R}$.
		In the steady state, $m$, $\dot{V}$, $\dot{P}$, $\dot{\psi}$, and $\ddot{\psi}$ are zero.
		Then, both sides of \eqref{eq:P} become zero, and thus parameters $k_1$ and $k_2$ do not affect the steady-state characteristics.
		The mass flow rate equation \eqref{eq:m} becomes $\alpha m_{\rm in} = (1-\alpha) m_{\rm out}$.
		Therefore, $\bar{P}$ is obtained from $u$ irrespective of $A_0$.
		The equations of the seesaw motion \eqref{eq:ssem2} and the friction model \eqref{eq:ff} can be rewritten with $\tau-T_f = 0$ and
		\begin{eqnarray*}
			T_f = \begin{cases}
				\dfrac{(T_s+T_p){\rm sgn}(\tau) + c_s Z \tau}{1 + Zc_s},& {\sf if} \,\ \left| \tau \right| > T_s+T_p, \\
				\tau, & {\sf if}\,\ \left| \tau \right| \leq T_s+T_p.
			\end{cases}
			\label{eq:ff_0}
		\end{eqnarray*}
		Therefore, the parameters $T_p'$ and $\mu_s$ affect the steady-state characteristics.
		Considering the above-described analysis, the estimated parameters are divided into two groups: transient and steady state.
\end{IEEEproof}

\begin{table}[tb]
	\centering
	\caption{Parameters of Antagonistic PAM System}
	\scriptsize
	\begin{tabular}[t]{llc} 
		\hline
		\hline
		$r_p$ & : radius of shaft (m) & \ \\
		$r$ & : radius of seesaw (m) &\ \\
		$L_0$ & : initial length of PAM (m) &\ \\
		$M$ & : weight of seesaw (kg) &\ \\
		$g$ & : gravitational acceleration (m/s$^2$) &\ \\
		$P_{\rm tank}$ & : source absolute pressure (Pa) &\ \\
		$P_{\rm out}$ & : atmospheric pressure (Pa) &\ \\
		$k$ & : specific heat ratio for air (--) &\ \\
		$R$ & : ideal gas constant (J/kg$\cdot$K) &\ \\
		$T$ & : absolute temperature (K) &\ \\
		$J$ & : moment of inertia of seesaw (kg$\cdot$m$^2$) &\ \\
		$k_s$ & : coefficient of static torque of seesaw (N$\cdot$m/rad) &\ \\
		$c_s$ & : viscous friction coefficient (N$\cdot$s) &\ 
		\raisebox{6.5mm}[0pt][0pt]{\hbox{\rotatebox{90}{Directly measurable}}}\\
		\hline
		$D_1,  D_2,  D_3$ & : coefficients of polynomial (m, m$^2$, m$^3$) &\ \\
		$p_{v1i}$, $p_{v2i}$, $p_{w1i}$, $p_{w2i}$  & : coefficient of force for PAM$i$ (--) &\ \\ 
		$A_{1i}$, $A_{2i}$ & : orifice area of PDCV$i$ (m$^2$) &\ \\
		$k_1,  k_2$ & : polytropic indexes (--) &\ \\
		$T_p'$ & : Coulomb friction coefficient of PAM (--) &\ \\
		$\mu_s$ & : Coulomb friction coefficient of shaft (--) &\ 
		\raisebox{3.0mm}[0pt][0pt]{\hbox{\rotatebox{90}{Estimated}}}\\
		\hline
		\hline
	\end{tabular}
	\label{tbl:variables}
\end{table}

	\begin{table}[t]
		\centering
		\caption{Identified Parameters of Antagonistic PAM System}
		\scriptsize
		\begin{tabular}{lrlr}
		\hline
		\hline
		$r_p$ (m) & 0.006 & $D_1$ (m) & $-2.440 \times 10^{-2}$\\
		$r$ (m) & 0.0365 & $D_2$ (m$^2$) & $6.824 \times 10^{-3}$\\
		$L_0$ (m) & $0.165$ & $D_3$ (m$^3$) & $-4.254 \times 10^{-4}$\\
		$M$ (kg) & $0.256$ & $p_{v11}$ (--)& $7.045 \times 10^{-3}$\\
		$g$ (m/s$^2$) & $9.80$ & $p_{v21}$ (--)& $-1.017 \times 10^{-3}$\\
		$P_{\rm tank}$ (Pa) & $0.7100\times 10^{6}$ & $p_{w11}$ (--)& $-5.568 \times 10^{2}$\\
		$P_{\rm out}$ (Pa) & $0.1013 \times 10^{6}$ &$p_{w21}$ (--)& $72.86$\\
		$k$ (--) & $1.40$ & $p_{v12}$ (--)& $6.423 \times 10^{-3}$\\
		$R$ (J/kg$\cdot$K) & $287$ & $p_{v22}$ (--)& $-9.184 \times 10^{-4}$\\
		$T$ (K) & $293$ & $p_{w12}$ (--)& $-197.8$\\
		$J$ (kg$\cdot$m$^2$) & $4.263 \times 10^{-4}$ & $p_{w22}$ (--)& $-15.75$\\
		$k_s$ (N$\cdot$m/rad) & $4.117 \times 10^{-4}$ & $A_{11}$ (m$^2$) & 5.184 $\times 10^{-8}$\\
		$c_s$ (N$\cdot$s) & $2.256 \times 10^{-3}$ & $A_{12}$ (m$^2$) & 7.776 $\times 10^{-8}$\\
		$k_1$ (--) & 1.100 & $T_p'$ (--) & $4 \times 10^{8}$\\
		$k_2$ (--) & 0.4545 & $ \mu_s$ (--) & 0.2\\
		\hline
		\hline
		\end{tabular}
		\label{tbl:parameters2pam}
	\end{table}
	
\begin{figure}[ht]
	\centering
	\subfloat[Simulation result of static relation between open rate and pressure.]{\includegraphics[width=0.3\hsize]{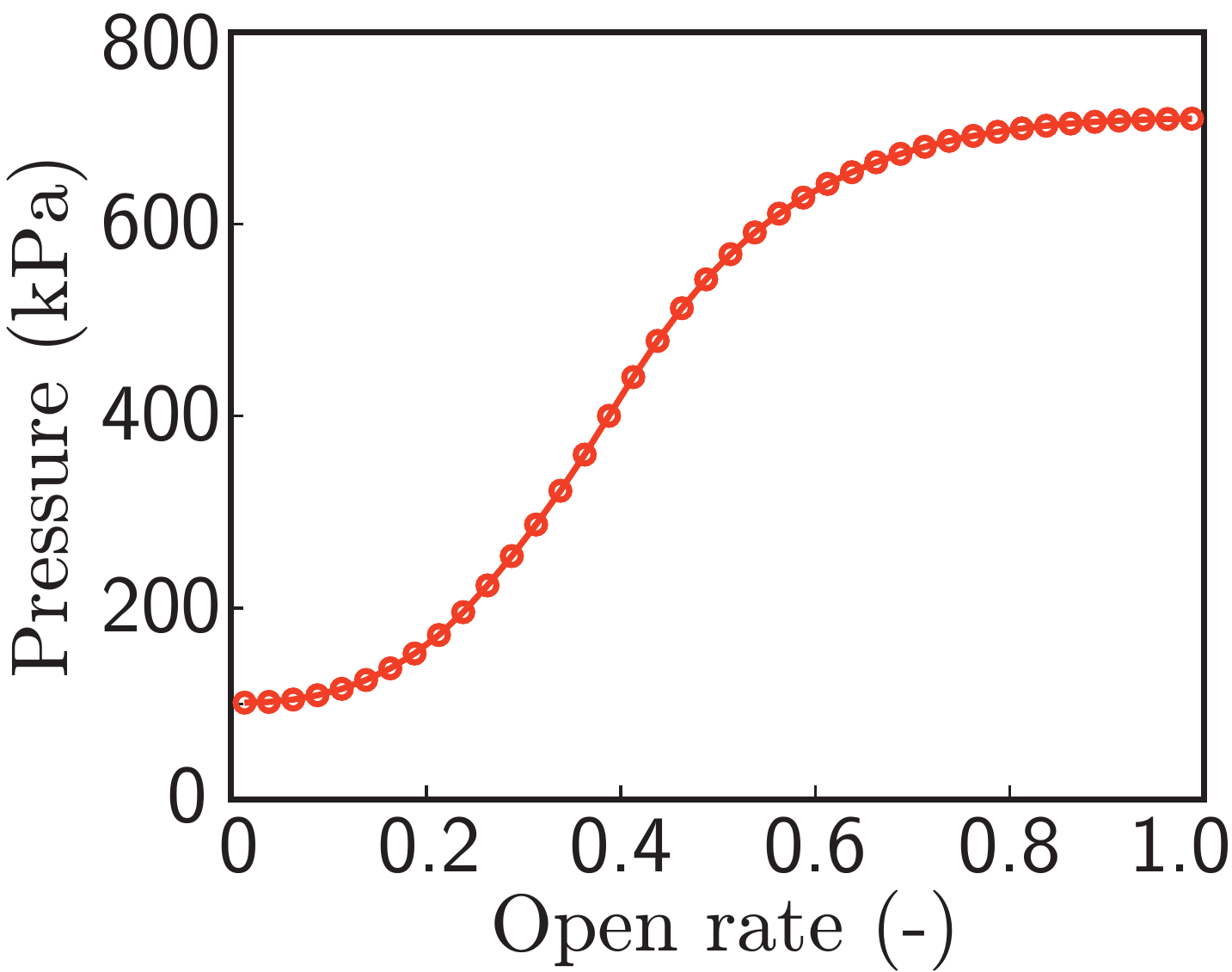}
	\label{subfig:a}}
	\hspace{0.5mm}
	\subfloat[Experimental result of static relation between input voltage and pressure.]{\includegraphics[width=0.3\hsize]{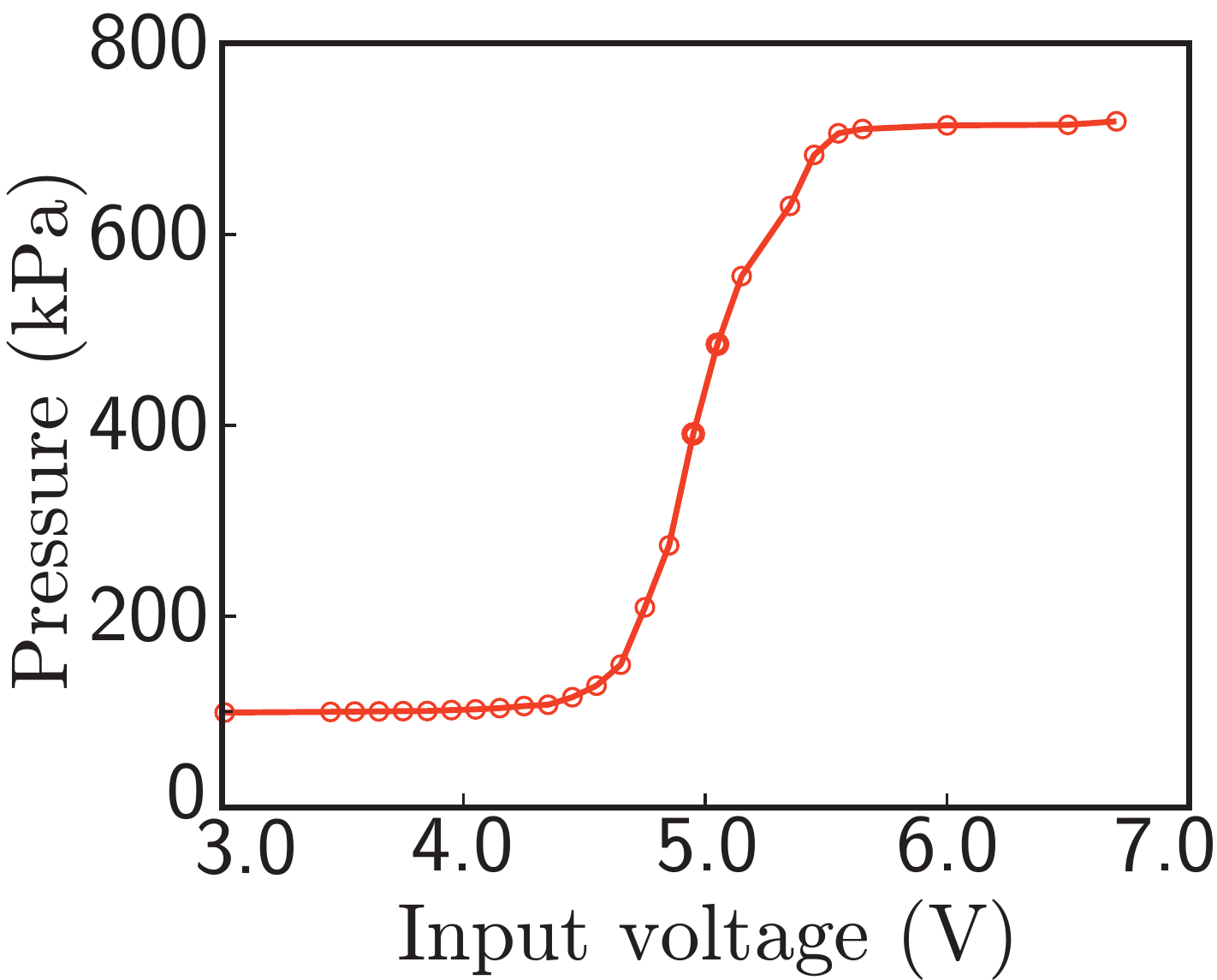}%
	\label{subfig:b}}
	\hspace{0.5mm}
	\subfloat[Static relation between input voltage and open rate.]{\includegraphics[width=0.3\hsize]{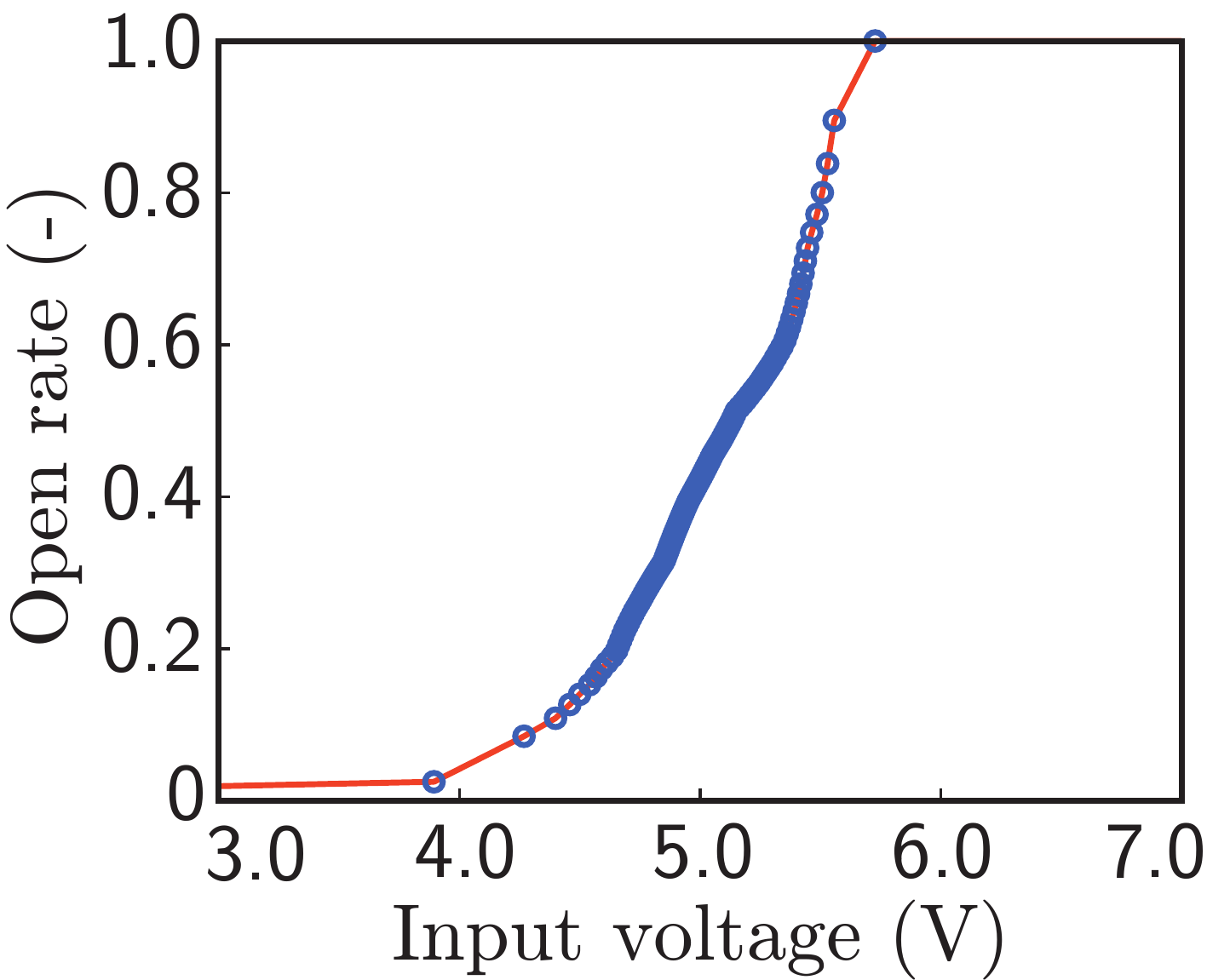}%
	\label{subfig:c}}
	\caption{A procedure for obtaining the function $\kappa(u)$.}
	\label{fig:get_kaido}
\end{figure}

Next, a three-step method is used for obtaining the function $\alpha=\kappa(u)$ appearing in Section \ref{subsec:FD}.
First, we obtain the relationship between the given open rate and the steady-state pressure through numerical simulations with the proposed model (Fig.\ \ref{fig:get_kaido}\subref{subfig:a}).
Second, we obtain the relationship between the input voltage to the PDCV and the experimentally obtained steady pressure (Fig.\ \ref{fig:get_kaido}\subref{subfig:b}).
Finally, by using the obtained relationships, we compare the steady pressure and obtain the relationship between the input voltage and the open rate, indicated by blue circles in Fig.\ \ref{fig:get_kaido}\subref{subfig:c}.
Then, a linear interpolation of these points is conducted to obtain the red solid line in Fig.\ \ref{fig:get_kaido}\subref{subfig:c}.

\begin{figure*}[ht]
	\centering
	\subfloat[Influence of changing~$T_p'$.]{\includegraphics[width=0.4\hsize]{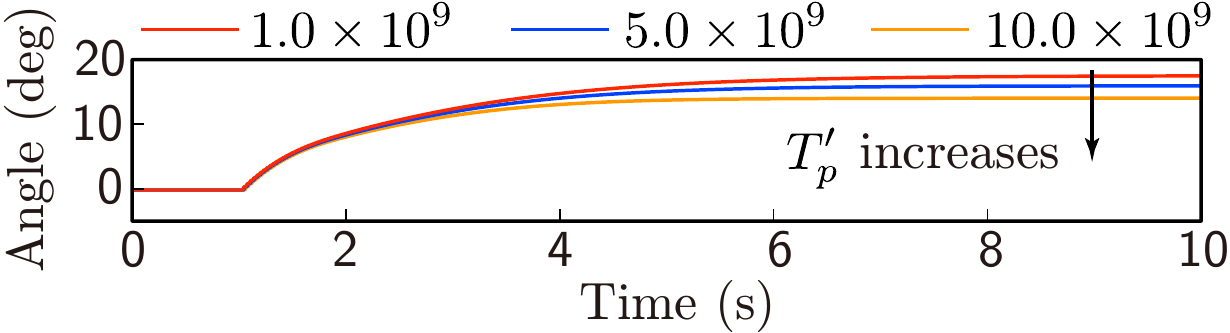}
	\label{subfig:para_Tp}}
	\hspace{2mm}
	\subfloat[Influence of changing $\mu_s$.]{\includegraphics[width=0.4\hsize]{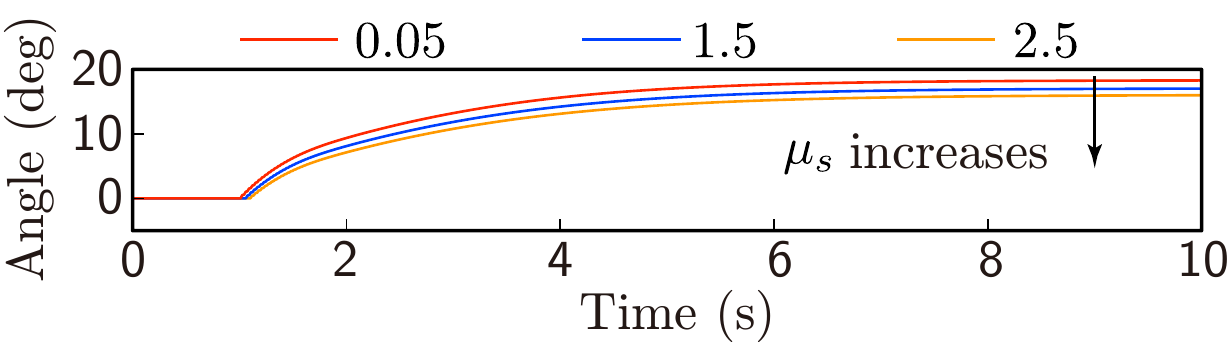}
	\label{subfig:para_mu}}
	\hspace{2mm}
	\subfloat[Influence of changing $A_0$.]{\includegraphics[width=0.4\hsize]{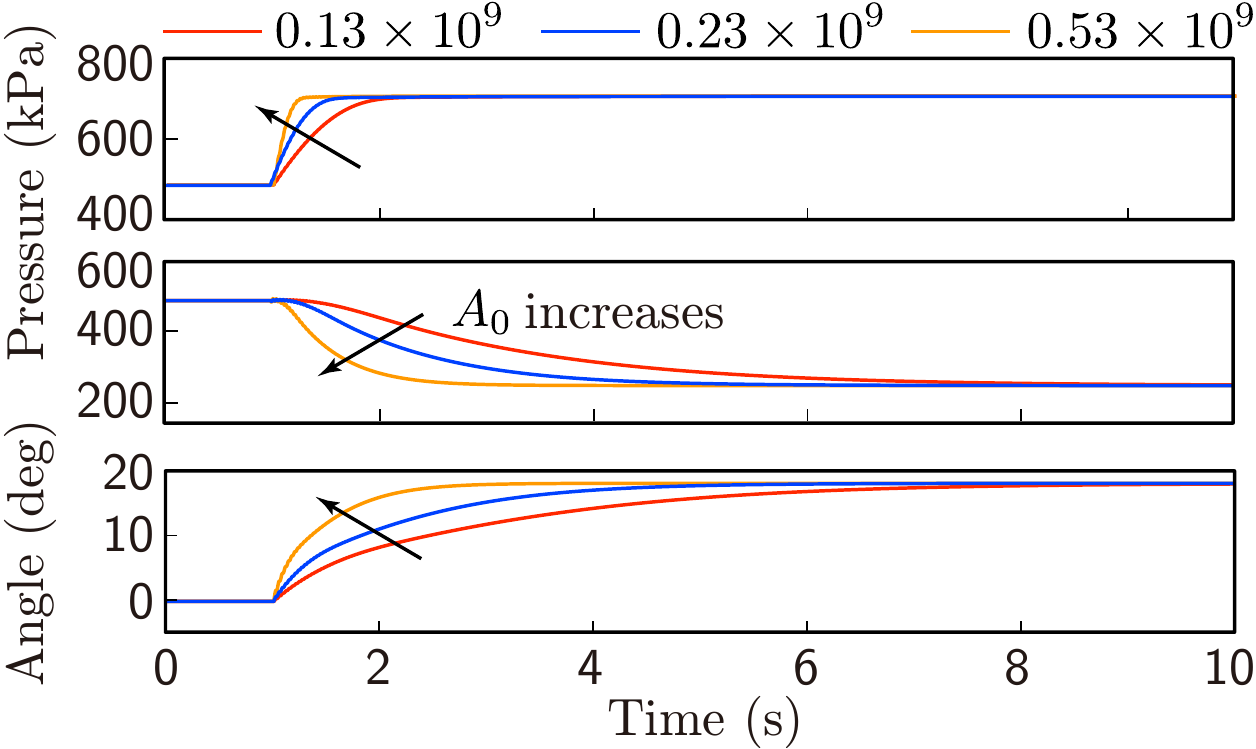}
	\label{subfig:para_A}}
	\hspace{2mm}
	\subfloat[Influence of changing $k_1$ and $k_2$.]{\includegraphics[width=0.4\hsize]{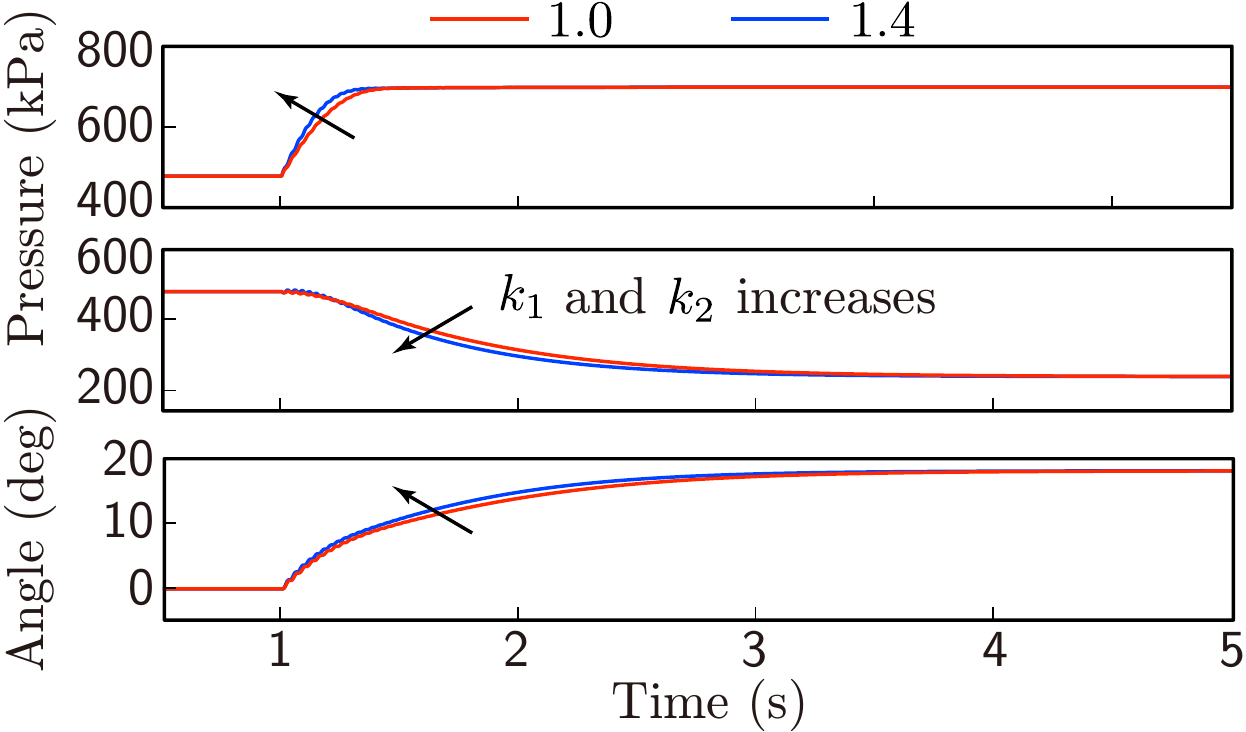}
	\label{subfig:para_k}}
	\caption{Influence of changing parameters.}
	\label{fig:change_para}
\end{figure*}

Based on \textbf{Proposition\ \ref{prop1}}, this study provides a time-domain parameter identification procedure for the proposed model.
In this procedure, first, we determine the directly measurable parameters. Second, we identify $D_1,~D_2$, and $D_3$ using a graduated cylinder \cite{2012_Itto}.
Third, we identify $p_{v1i}$, $p_{v2i}$, $p_{w1i}$, and $p_{w2i}$ by using measurement data and \eqref{eq:force}.
Then, we determine the parameters $T_p'$ and $\mu_s$, which affect the steady-state characteristics of the see saw, such that the steady-state error between the measurement and the simulation data of the joint angle becomes relatively small through trial-and-error.
Finally, we determine the parameters $A_0$, $k_1$, and $k_2$, which affect transient characteristics, such that the transient response error between the measurement and the simulation data becomes relatively small. The resulting model parameters are listed in Table\ \ref{tbl:parameters2pam}, where $i\in{\mathcal I}$ denotes the number of PAMs.
In addition, we provide a guideline that as $T_p'$ or $\mu_s$ increases, the steady-state value of the joint angle decreases, and that as $A_0$, $k_1$, or $k_2$ increases, the transient response becomes faster.
This guideline helps determine the parameters, and it is illustrated in Fig. \ref{fig:change_para}.

\section{UKF-based Angle and Torque Estimators}\label{sec:ModelValidation}
This study uses the UKF, proposed in \cite{ukf_2000}, to estimate the state of the antagonistic PAM actuator system.
The antagonistic PAM system used in this study contains a pressure sensor, a rotary encoder, and a torque meter.
The torque meter and rotary encoder are installed on the joint part of the devices for performing measurements.
However, this increases the device weight and load acting on patients wearing real assistant devices.
The pressure sensor need not be set up at the actuator joint part.
If the joint angle and torque can be estimated, the time required for designing and constructing a PAM actuator system could be further reduced.
This study uses the information observed by only the pressure sensor for the UKF.

\begin{figure}[t]	
		\centering
		\includegraphics[width=2.7in]{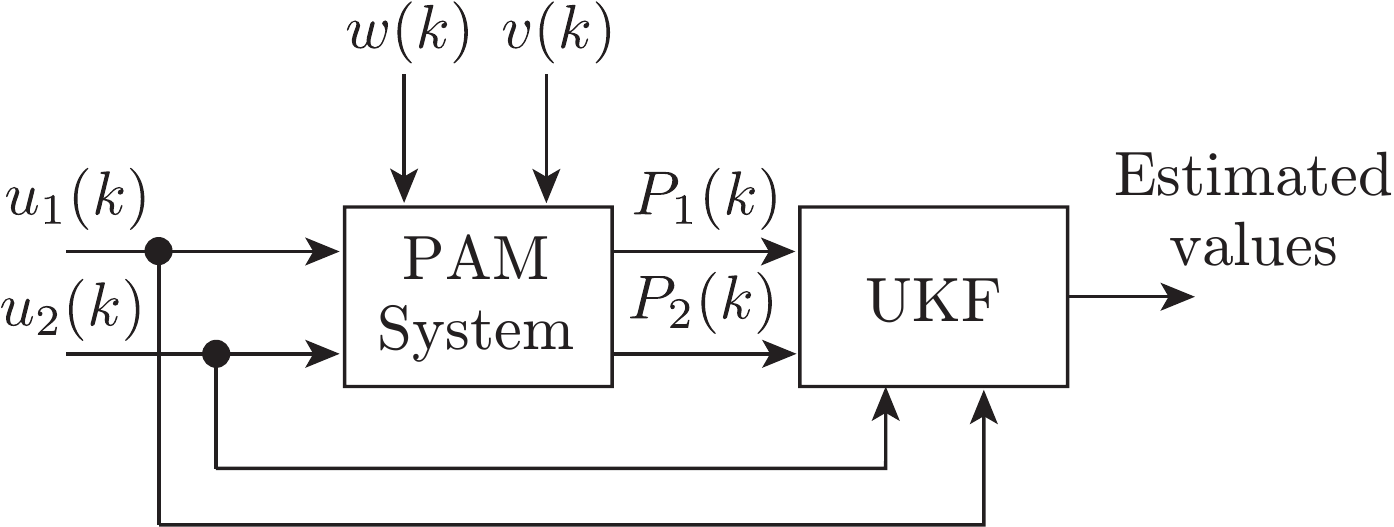}
		\caption{UKF-based torque estimation}
		\label{fig:block_estimation}
	\end{figure}

The UKF needs a discrete-time model of the system, and two noise signals are assumed to be added to the system:
	\begin{subequations}
		\begin{align*}
			x(k+1) &= f_\sigma(x(k), u(k)) + v(k) \hspace*{1ex}{\sf if} \ x(k) \in \mathcal{X}_{\sigma},\hspace*{2ex} \\
			y(k) &= g(x(k)) + w(k),
		\end{align*}
		\label{eq:pam_ukf}
	\end{subequations}
where
	\begin{align*}
		g(x(k)) = \left[\begin{matrix}0 & 0 & 1 & 0 \\ 0 & 0 & 0 & 1 \end{matrix}\right]x(k).
	\end{align*}
The discrete representation of \eqref{eq:pam} is obtained using the fourth-order Runge-Kutta method.
$v\in \mathbb{R}^4$ and $w\in \mathbb{R}^2$ are the process and observation noise in the system, respectively; 
specifically, these are a zero-mean white noise with the covariance matrix $Q\in \mathbb{R}^{4\times 4}$ and $R\in \mathbb{R}^{2 \times 2}$, respectively.
Fig. \ref{fig:block_estimation} shows a block diagram of the state estimator.
The estimated variables are the state values $\hat{P_1}, \hat{P_2}$, and $\hat{\psi}$; 
the contracting force $\hat{F_1}$ and $\hat{F_2}$; and the torque $\hat{\tau}$.
$\hat{F_1}$ and $\hat{F_2}$ are obtained from \eqref{eq:force} with the estimated state values.
$\hat{\tau}$ is obtained from \eqref{eq:tauPAM} using $\hat{F_1}$ and $\hat{F_2}$.
The UKF assumes the process noise and observation noise in a system to be regular white noises.
This assumption simplifies the calculation of the probability distribution. 
The UKF can remove these noises and estimate unknown variables by updating the covariance matrix and the mean of the probability distribution of the state value.

\begin{figure}[t]
		\centering
		\includegraphics[width=2.7in]{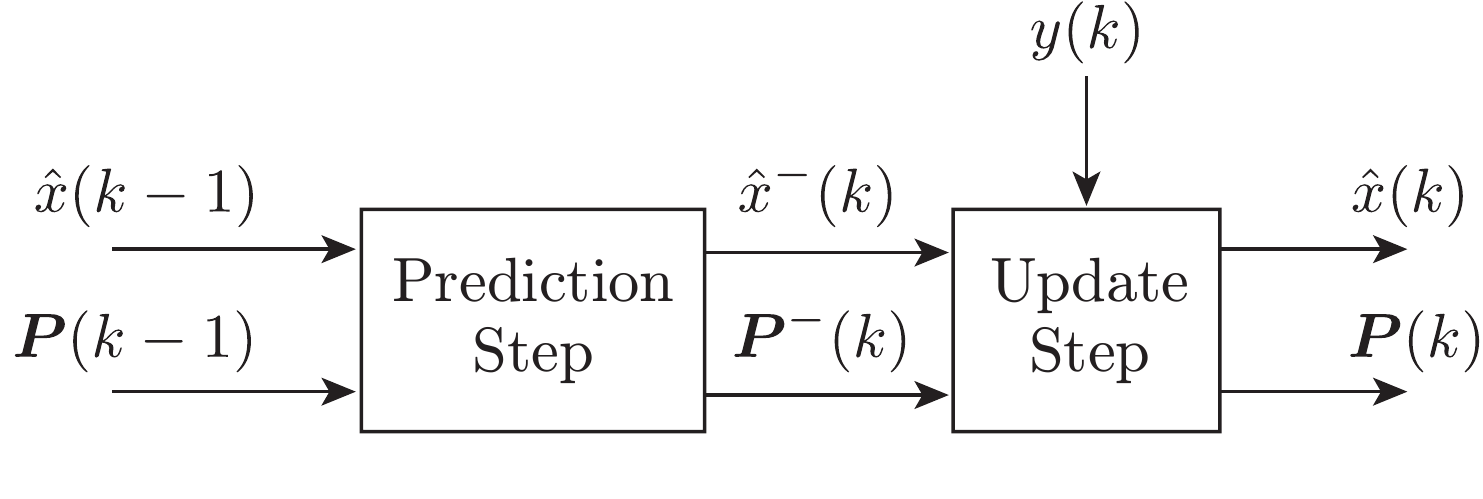}
		\caption{Algorithm of unscented Kalman filter}
		\label{fig:ukf_block}
	\end{figure}

The UKF algorithm involves two steps: prediction and update.
In the prediction step, the state at the current time step is estimated using the estimated state at the previous time; 
the estimated state in the prediction step is called a priori state estimate because it is calculated using previous information.
In the update step, the estimated states are refined using observation information; the updated state is called a posteriori estimate.
Fig. \ref{fig:ukf_block} shows the block diagram of the UKF.
The detailed algorithm of the UKF is described below.

\subsection{Prediction Step}
In the prediction step, the priori state $\hat{x}^-(k)$ is estimated from the estimated state of the previous time step $\hat{x}(k-1)$.
$\hat{x}^-(k)$ is obtained from the probability distribution of the state vector.
The UKF calculates a probability distribution after nonlinear transformation by using some sample points.
These sample points are called sigma points, and the sigma points at step $k$, $\mathcal{X}_i(k-1)$, are chosen as follows:
	\begin{align}
		\mathcal{X}_0(k-1)&=\hat{x}(k-1), \\
		\mathcal{X}_i(k-1)&=\hat{x}(k-1) + (\sqrt{(n+\kappa)\bm{P}(k-1)})_i,\\
		\mathcal{X}_{i+n}(k-1)&=\hat{x}(k-1) - (\sqrt{(n+\kappa)\bm{P}(k-1)})_i,
	\end{align}
where $i=0,1,2 \dots, n$ is the number of sigma points, $n = 4$ is the dimension of the state vector, and $(\sqrt{(n+\kappa)\bm{P}(k)})_i$ is the $i$th row of the square root matrix for $(n+\kappa)\bm{P}(k)$.
Next, the sigma points $\mathcal{X}_i(k)$ are updated using the state equation $f_{\sigma}$ as follows:
\footnotesize
	\begin{align}
		\mathcal{X}^-_i(k) = f_{\sigma}(\mathcal{X}_i(k-1), u(k-1))),\ \ i=0,1,2 \dots 2n.
	\end{align}
\normalsize
A priori state estimate and a covariance matrix are calculated as follows:
\footnotesize
	\begin{align}
		\hat{x}^-(k)&=\sum_{i=0}^{2n}W_i \mathcal{X}^-_i{(k)},\\
		\bm{P}^-(k)&=\sum_{i=0}^{2n}W_i(\mathcal{X}^-_i{(k)}-\hat{x}^-(k))\cdot(\mathcal{X}^-_i(k)-\hat{x}^-(k))^T + Q,
	\end{align}
\normalsize
\!\!where $W_0=\kappa/(n+\kappa)$ and $W_i=1/2(n+\kappa)$\ ($i = 1,2,...,2n$), where $\kappa$ is the scaling parameter.
The sigma points $\mathcal{X}^-_i(k)$ are recalculated using $\hat{x}^-(k)$ and $\bm{P}^-(k)$, and the sigma points are converted using $h$ as follows:
	\begin{align}
		\mathcal{Y}^-_i(k) = g(\mathcal{X}^-_i(k)).
	\end{align}
The observed value and covariance matrix are calculated as follows:
\footnotesize
	\begin{align}
		\hat{y}^-(k)&=\sum_{i=0}^{2n}W_i \mathcal{Y}^-_i{(k)},\\
		\bm{P^-_{yy}}(k)&=\sum_{i=0}^{2n}W_i(\mathcal{Y}^-_i{(k)} - \hat{y}^-(k))\cdot(\mathcal{Y}^-_i(k)-\hat{y}^-(k))^T + R,\\
		\bm{P^-_{xy}}(k)&=\sum_{i=0}^{2n}W_i(\mathcal{X}^-_i{(k)} - \bm{\hat{x}^-}(k))\cdot(\mathcal{Y}^-_i(k)-\hat{y}^-(k))^T.
	\end{align}
\normalsize
\subsection{Update Step}
In the update step, the estimated state is refined using the observed information $y(k)$ and the modified state at the current time step $\hat{x}(k)$ is calculated.
A posteriori estimate and a covariance matrix are calculated as follows:
	\begin{align}
		{\hat x}(k)&=\hat{x}^-(k) + K(k)(y(k) - \hat{y}^-(k)),\\
		\bm P(k)&=\bm P^-(k) - K(k)\bm {P^-_{yy}}(k)K(k)^T,
	\end{align}
where $K(k) = \bm P^-_{xy}(k)/(P^-_{yy}(k) + R)$.
A UKF algorithm for estimating the state is summarized in {\bf Algorithm~1}.

\begin{algorithm}[t]
\caption{Unscented Kalman filter}\label{alg:UKF}
\begin{algorithmic}[1] 
\Function{UKF}{$\hat{x}(k-1),~\bm{P}(k-1),~u(k-1),~y(k)$}
\State $\triangleright$ Prediction step

\State \textbf{for}~$i=1$~\textbf{to}~$n$
\State ~~~~Calculate (13) to (15)
\State \textbf{endfor}

\State \textbf{for}~$i=1$~\textbf{to}~$2n$
\State ~~~~Calculate (16)
\State \textbf{endfor}

\State Calculate (17) and (18)

\State \textbf{for}~$i=1$~\textbf{to}~$n$
\State ~~~~Calculate (19)
\State \textbf{endfor}

\State Calculate (20) to (22)

\Statex
\State $\triangleright$ Update step

\State Calculate (23) and (24)
	
\State \textbf{return} ${\hat x}(k),~\bm{P}(k)$
\EndFunction
\end{algorithmic}
\end{algorithm}
\normalsize

\section{Model Validation}\label{sec:CS}
The model validation procedure consists of three steps:
(1) comparison of the angle and torque estimations in the offline simulation, as described in Section \ref{subsec:OfflineSimu};
(2) implementation and confirmation of the possibility of real-time estimation with the UKF, as described in Section \ref{sec:ole}; and 
(3) evaluation of whether the proposed model helps to design and configure sensor-less control systems.
In addition, the root mean square error (RMSE), an $\ell_\infty$-norm of the error, and the maximum estimation ratio to a sensor signal $\xi(k)$,
\begin{align}
	\mathbb{I}(z,N) &:=\sqrt{\dfrac{1}{N}\sum_{k=0}^{N} z^2(k)}, \quad z\in\mathbb{R},\ N\in\mathbb{Z}_{\geq 0},\nonumber\\[2ex]
	\mathbb{M}(z) &:=||z||_\infty =\max_{k\in\mathbb{Z}_{\geq 0}} |z(k)|, \nonumber\\[2ex]
	\mathbb{Q}(z,\xi,N) &:=\frac{\mathbb{M}(z)}{\max_{k=0}^{N}{\xi(k)}-\min_{k=0}^{N}{\xi(k)}}, \nonumber
\end{align}
are introduced to quantify the estimation accuracy through $N=130\times 10^3$ steps, as described below.
The UKF parameters are determined as $\bm{P}(0)={\rm diag}(10^{-5},10^{-4},\ 10^{6},\ 10^{6})$, $R={\rm diag}(10^8,10^8)$, $Q={\rm diag}(10^{-5},10^{-4},\ 10^{6},\ 10^{6})$, and $\kappa=0$.

\subsection{Offline Estimation}\label{subsec:OfflineSimu}
\begin{figure}[t]
		\centering
		\includegraphics[width=0.95\hsize]{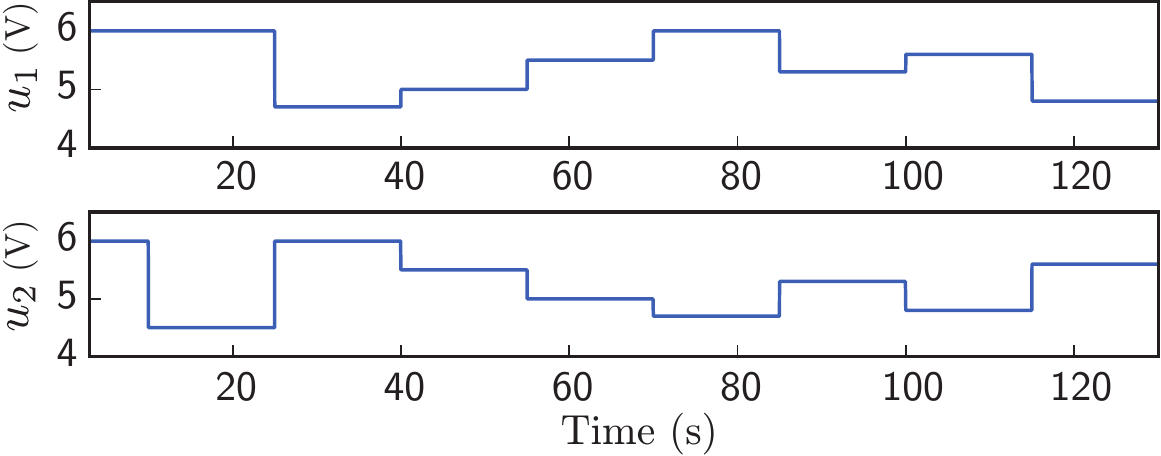}
		\caption{Control input profiles to PDCV1 (upper) and PDCV2 (lower) considered in the offline and online estimations.}
		\label{fig:input_vol}
	\end{figure}

A numerical simulation is conducted to qualitatively demonstrate that the model-based UKF is better than the model alone in terms of the estimation of state information. 
This study considers the following two methods.
The estimation-by-model method computes (estimates) the time response of the state variable from the proposed model, and the estimation-by-UKF method with the proposed model obtains estimates of the state variable by using the control input signals and the experimentally measured output signals.
In addition, the simulation considers the common control inputs shown in Fig.\ \ref{fig:input_vol} that cover a wide pressure range of 200--700~kPA.

\subsubsection{Joint angle estimation}
\begin{figure}[t]
		\centering
		\subfloat[Time responses of inner pressure of PAM1 (upper) and PAM2 (lower).]
		{\includegraphics[width=1.0\hsize]{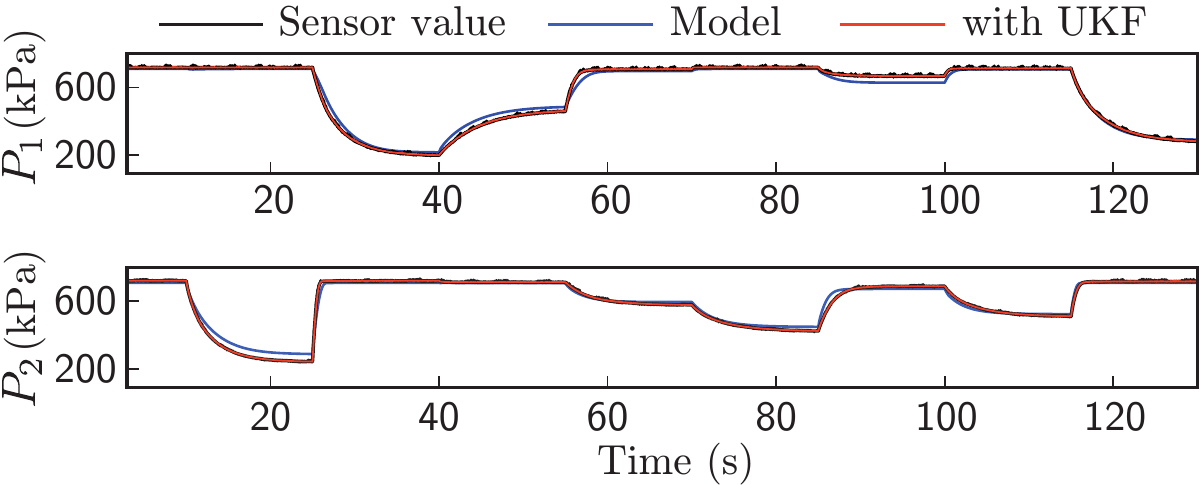}
		\label{subfig:angle_off_state}}\\
		\subfloat[Time responses of measured and estimated joint angles (upper) and their estimation errors $\zeta_\psi$~(lower).]
		{\includegraphics[width=1.0\hsize]{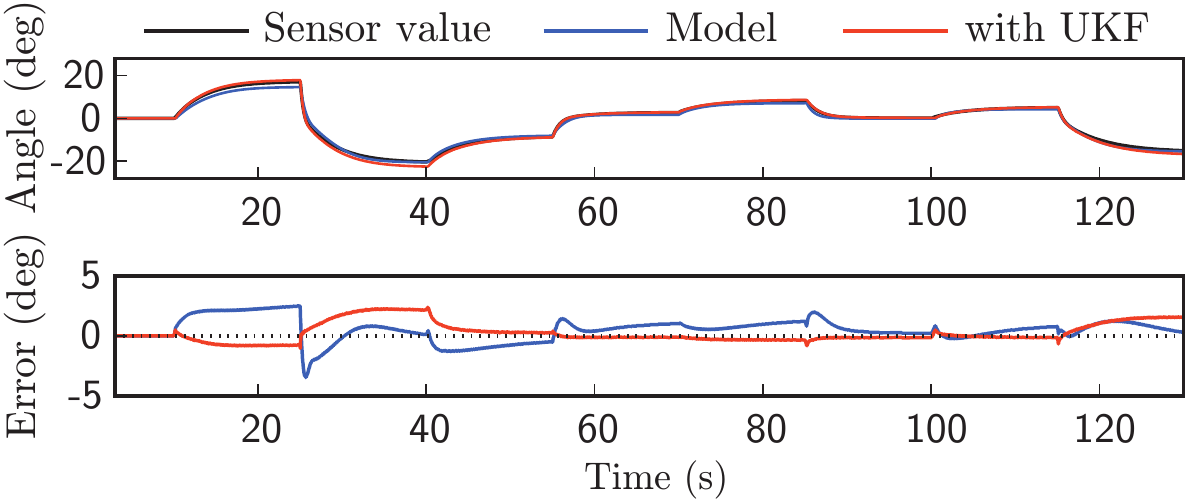}
		\label{subfig:angle_off_esti}}
		\caption{Comparison of offline estimations of the joint angle in the simulation.}
		\label{fig:angle_off}
	\end{figure}

Fig.\,\ref{fig:angle_off} shows a simulation result of the pressure, joint angle, and its estimation error. 
The black line indicates the value measured by the sensor; blue line, the value estimated by the model; and red line, the value estimated by the UKF. 
As shown in Fig.\ \ref{fig:angle_off}\subref{subfig:angle_off_state}, although there is a deviation in $P_2$ from the sensor signal at around 20 s, both pressure estimations are almost the same as the signal measured by the sensor.
Fig.\ \ref{fig:angle_off}\subref{subfig:angle_off_esti} shows comparisons of the responses of the joint angle and its estimation error. 
The RMSEs with the proposed model alone and the UKF-based estimation are respectively $\mathbb{I}(z_\psi,N)=1.053$ and $\mathbb{I}(z_\psi,N)=0.838$, and 
the $\ell_\infty$-norm values with the proposed model alone and the UKF-based estimation are respectively $\mathbb{M}(z_\psi)=3.50$$^\circ$ and $\mathbb{M}(z_\psi)=2.42$$^\circ$, where $z_\psi:=\psi-\hat\psi$.
The ratio of $\mathbb{M}(z_\psi)$ to a range of used signals is $\max \psi(k)=19.1$ and $\min \psi(k)=-20.3$, that is, 
$\mathbb{Q}(z_\psi,\psi,N)=0.0887$ (8.87 \%) with the proposed model alone and
$\mathbb{Q}(z_\psi,\psi,N)=0.0613$ (6.13 \%) with the UKF-based estimation.
These smaller values indicate that the model-based UKF better estimates the joint angle.

\subsubsection{Torque estimation}
\begin{figure}[t]
		\centering
		\subfloat[Responses of inner pressure of PAM1 (upper) and PAM2 (lower).]
		{\includegraphics[width=1.0\hsize]{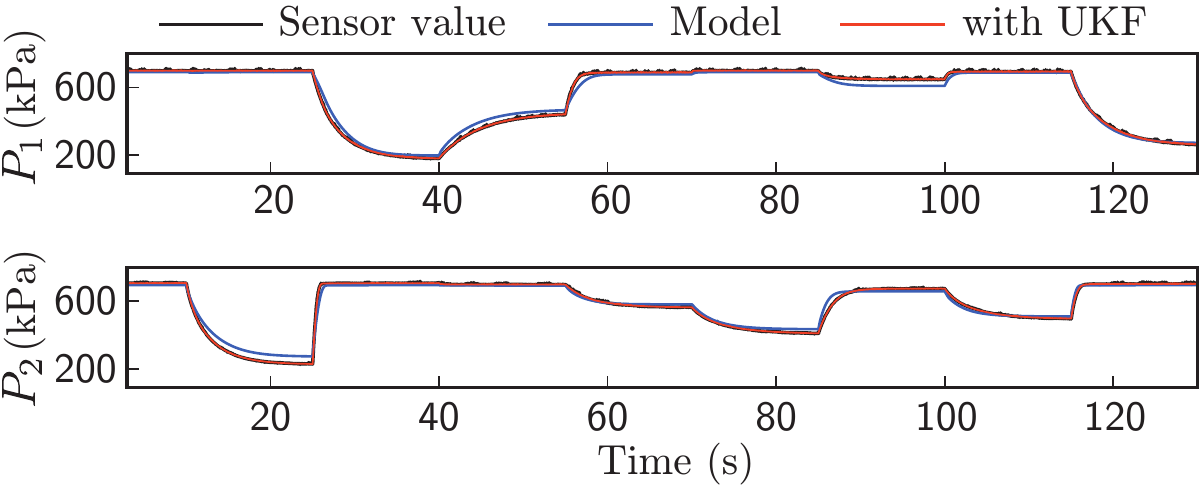}
		\label{subfig:torque_off_state}}\\
		\subfloat[Torque responses (upper) and estimation errors $\zeta_\tau$~(lower).]
		{\includegraphics[width=1.0\hsize]{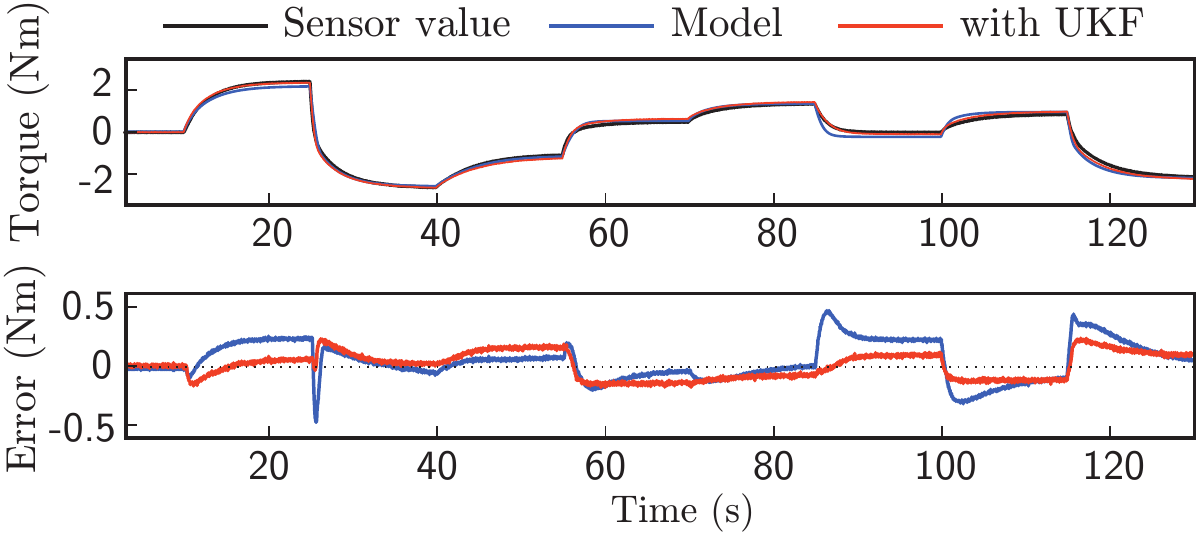}%
		\label{subfig:torque_off_esti}}
		\caption{Comparison of offline estimations of the torque in the simulation.}
		\label{fig:torque_off}
	\end{figure}

For torque measurements, the joint angle was fixed at 0$^\circ$.
Fig.\ \ref{fig:torque_off} shows a simulation result of the pressure, the torque, and its estimation error. 
The line colors in this figure have the same meaning as those in Fig.\,\ref{fig:angle_off}.
Fig.\ \ref{fig:torque_off}\subref{subfig:torque_off_state} shows that both pressure estimations are almost the same as the signal measured by the sensor.
Fig.\ \ref{fig:torque_off}\subref{subfig:torque_off_esti} shows comparisons of the responses of the torque and its estimation error. 
The RMSEs with the proposed model alone and the UKF-based estimation are respectively $\mathbb{I}(z_\tau,N)=0.162$ and $\mathbb{I}(z_\tau,N)=0.108$, and 
the $\ell_\infty$-norm values with the proposed model alone and the UKF-based estimation are respectively $\mathbb{M}(z_\tau)=0.494$ and $\mathbb{M}(z_\tau)=0.238$, where $z_\tau:=\tau-\hat\tau$.
The ratio of $\mathbb{M}(z_\tau)$ to a range of used signals is $\max \tau(k)=2.15$ and $\min \tau(k)=-2.68$, that is, 
$\mathbb{Q}(z_\tau,\tau,N)=0.1023$ (10.23 \%) with the proposed model alone and
$\mathbb{Q}(z_\tau,\tau,N)=0.0494$ (4.94 \%) with the UKF-based estimation.
These smaller values indicate that the model-based UKF better estimates the torque as well.

\subsection{Online Estimations}\label{sec:ole}
\begin{figure}[t]
	\centering
	\subfloat[Time responses of the joint angle and the estimation (upper) and its error $\zeta_\psi$~(lower).]
	{\includegraphics[width=1.0\hsize]{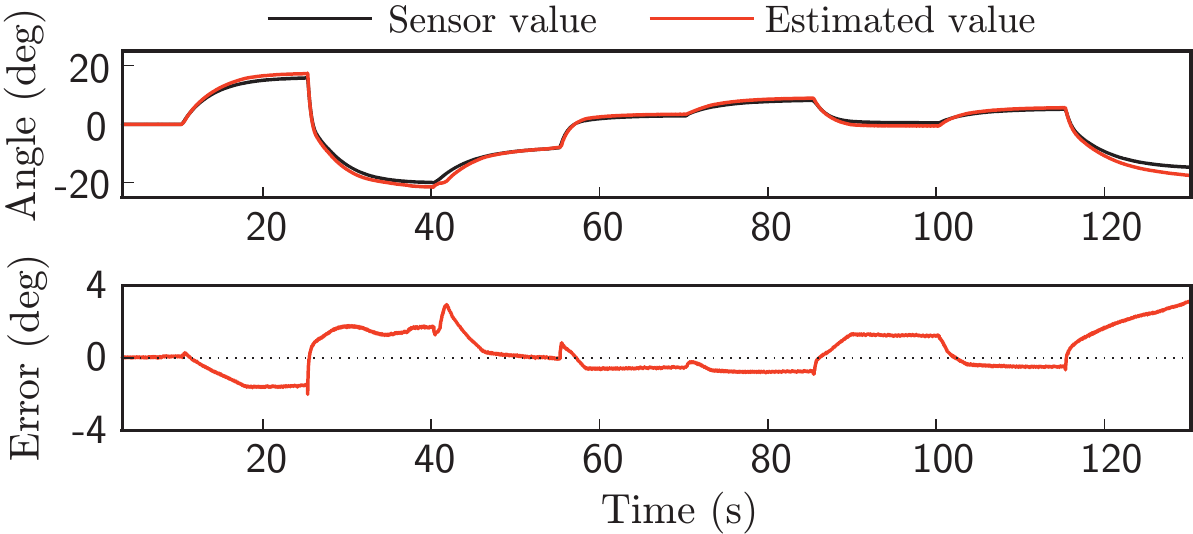}
	\label{subfig:angle_on_esti}}\\
	\subfloat[Time responses of the torque and the estimation (upper) and its error $\zeta_\tau$~(lower).]
	{\includegraphics[width=1.0\hsize]{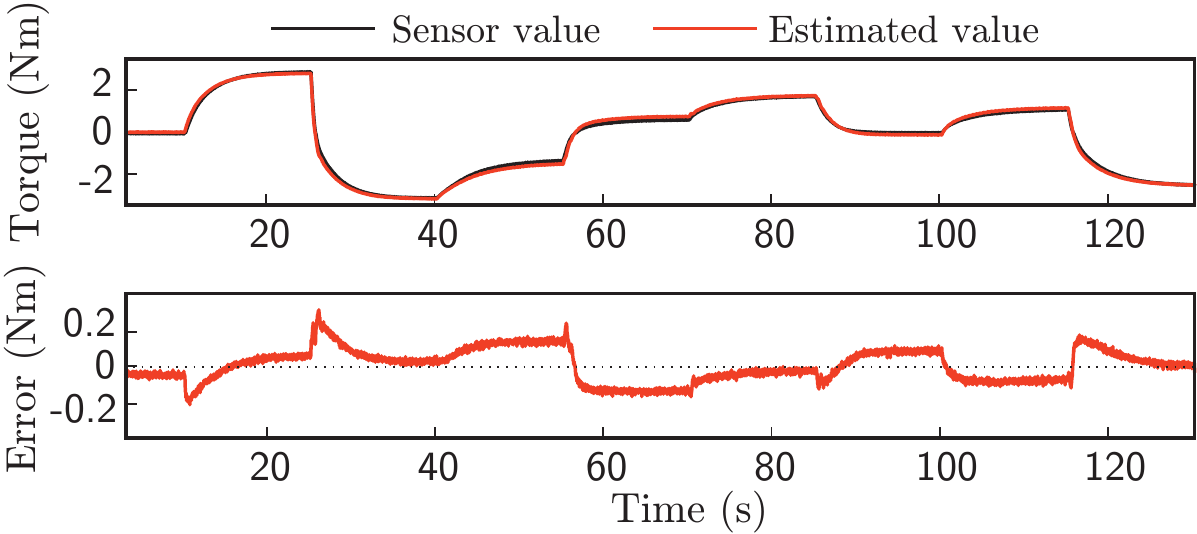}%
	\label{subfig:torque_on_esti}}
	\caption{Results of online estimation of the joint angle and torque in a practical setting.}
	\label{fig:online_esti}
\end{figure}

\begin{figure}[t]
	\centering
	\includegraphics[width=0.95\hsize]{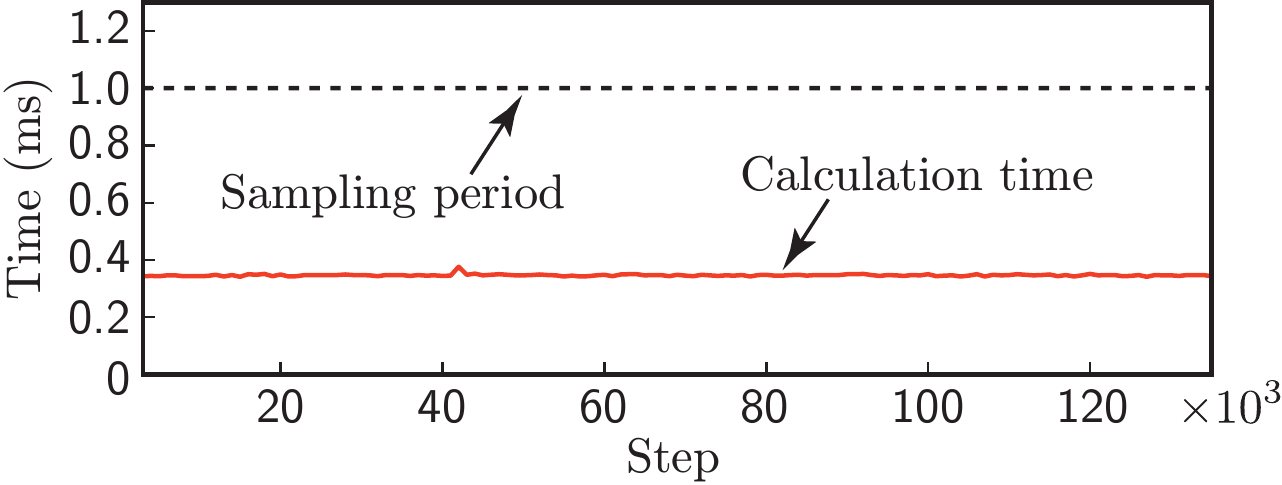}
	\caption{Processing time required to estimate the joint angle and torque; it is less than the sampling period of 1.0 ms.}
	\label{fig:ctime}
\end{figure}

The UKF-based estimation method is implemented in the computer of the antagonistic PAM system, and online estimation is conducted to show that this process can be performed in real time with a sampling period of 1.0 ms.
The control inputs are the same as those shown in Fig.\ \ref{fig:input_vol}, and the online estimation results and measured processing time are shown in Figs.\,\ref{fig:online_esti} and \ref{fig:ctime}, respectively.

Fig.\ \ref{fig:online_esti} shows experimental results of the joint angle and torque. The black line indicates a value measured by the sensor, and the red line indicates a value estimated by the UKF. 
The estimation accuracy of the joint angle in Fig.\,\ref{fig:online_esti}\subref{subfig:angle_on_esti} is $\mathbb{I}(z_\psi,N)=1.16$ and $\mathbb{M}(z_\psi)=3.06$.
The estimation accuracy of the torque in Fig.\ \ref{fig:online_esti}\subref{subfig:torque_on_esti} is $\mathbb{I}(z_\tau,N)=0.0899$ and $\mathbb{M}(z_\psi)=0.299$.
The maximum estimation ratio is $\mathbb{Q}(z_\psi,\psi,N)=0.0791$ (7.91 \%) with $\max \psi(k)=17.2$ and $\min \psi(k)=-21.5$, and $\mathbb{Q}(z_\tau,\tau,N)=0.0601$ (6.01 \%) with $\max \tau(k)=2.50$ and $\min \tau(k)=-2.89$.
These smaller values indicate that the model-based UKF provides relatively good estimation performance.
Further, Fig.\,\ref{fig:ctime} confirms that the processing time required to estimate the joint angle and torque using the UKF is within the sampling period; the average processing time is 0.346 ms.
Therefore, the online estimations confirm that the proposed nonlinear model is detailed enough to be used for real-time sensor-less control of the PAM actuator system with relatively good estimation accuracy of $\leq$7.91 \%.

To evaluate the estimation performance of the proposed model, this study compared it with other models in the online estimation.
As for the joint angle estimation, we used the following four models: (1) Linear model, (2) Simple friction model found in \cite{2000_cald, 2016_Andrikopoulos} $F_1 = \mathrm{sgn}(\dot{x})F_r$, (3) Tondu's friction model in \cite{2000_Tondu, 2013_Tondu}, $F_2 = \mathrm{sgn}(\dot{x})[(f_s-f_k)e^\frac{-\dot{x}}{\dot{x}_k}+f_k]$, and (4) our proposed model without the friction term \eqref{eq:ff}. Here, model (1) means that the whole of the proposed model is replaced with a linear model, and models (2) and (3) mean that only the friction term \eqref{eq:ff} in the proposed model is replaced with the respective friction model.
The model parameters were set to $F_r=13.7$, $f_s=0.146$, $f_k=0.03285$, and $\dot{x}_k=0.8$. As for the joint torque estimation, we used three empirical models: (1) Eq. (6) in \cite{2005_Hildebrandt}, (2) Eq. (2) in \cite{2010_Sarosi}, and (3) Eq. (3) in \cite{2012_polymodel}, instead of the contraction force \eqref{eq:force}.
Tables \ref{tbl:compare_online} and \ref{tbl:compare_tau_on} show the resulting estimation performances, where the best performances are emphasized.
The scores confirm that the proposed model has the best estimation performance.

\begin{table}[t]
	\centering
	\caption{Quantitative comparison in online angle estimation.}
	\begin{tabular}{l  c  c  c  c }
	\hline\hline
	Model&RSME~$\mathbb{I}$&$l_\infty$-norm~$\mathbb{M}$&$\mathbb{Q}$\\ \hline
	Proposed model &\bf{1.16}&\bf{3.06}&\bf{0.0791}\\
	Linear model &2.61&6.74&0.192\\
	Simple friction model \cite{2000_cald, 2016_Andrikopoulos}&1.36&3.39&0.0990\\
	Tondu's friction model \cite{2000_Tondu, 2013_Tondu}&1.43&3.38&0.0983\\
	No friction term&1.57&3.87&0.113\\
	\hline\hline
	\end{tabular}
	\label{tbl:compare_online}
\end{table}

\begin{table}[t]
	\centering
	\caption{Quantitative comparison in online torque estimation.}
	\begin{tabular}{l c  c  c  c }
	\hline\hline
	Model&RSME~$\mathbb{I}$&$l_\infty$-norm~$\mathbb{M}$&$\mathbb{Q}$\\ \hline
	Eq. (6)&\bf{0.0899}&\bf{0.299}&\bf{0.0601}\\
	Polynomial model 1 \cite{2005_Hildebrandt}&0.422&0.687&0.135\\
	Polynomial model 2 \cite{2010_Sarosi}&0.323&0.544&0.106\\
	Polynomial model 3 \cite{2012_polymodel}&0.550&0.985&0.193\\ \hline
	\hline
	\end{tabular}
	\label{tbl:compare_tau_on}
\end{table}

\subsection{Application to Sensor-less Control}
\begin{figure}[tb]
	\centering
	\subfloat[An encoder-less control system.]{\includegraphics[width=0.9\hsize]{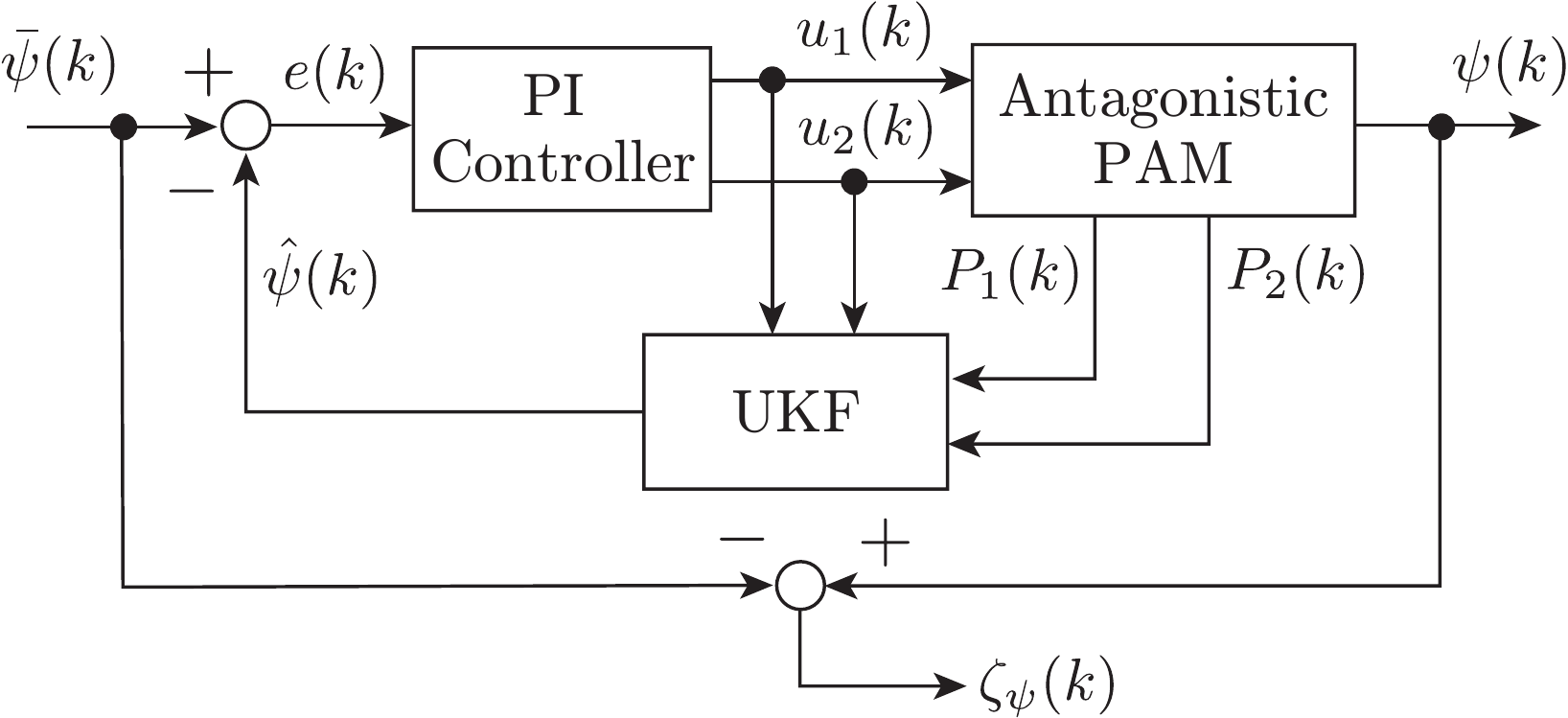}
	\label{subfig:angle_controller}}\\
	\subfloat[A torque-sensor-less control system.]{\includegraphics[width=0.9\hsize]{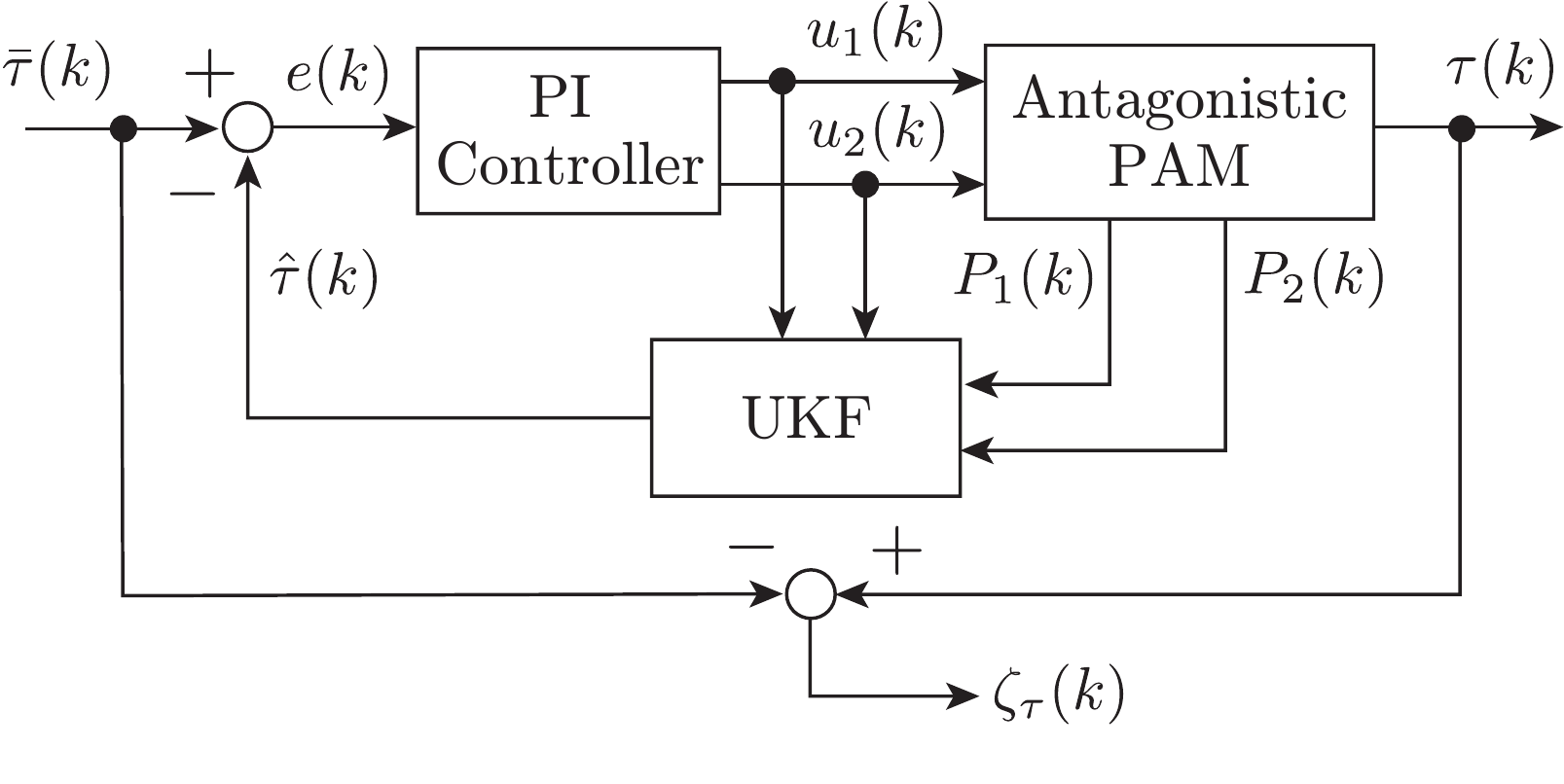}%
	\label{subfig:torque_controller}}
	\caption{Block diagrams of UKF-based sensor-less control system, where $\psi$ and $\tau$ are used only to evaluate the control performance.}
	\label{fig:sensorless_controllers}
\end{figure}

The UKF-based estimation is integrated into a practical antagonistic PAM control system, and the proposed nonlinear model is demonstrated to help in constructing sensor-less control systems.
The control object considered in the application is tracked to a given reference, and a PI control system is employed for this purpose.
Figs.\,\ref{fig:sensorless_controllers}\subref{subfig:angle_controller} and \subref{subfig:torque_controller} respectively show block diagrams of angle-sensor (encoder)-less and torque-sensor-less PAM control systems, with a feedback error $e$ between an estimation $\hat{\cdot}$ and a reference $\bar{\cdot}$.
The controller is written in the state-space representation as follows:
\begin{eqnarray*}
	\begin{cases}
		x(k+1) = x(k) + e(k),\\
	    u(k) = Cx(k) + De(k) + \beta,
	\end{cases}
\end{eqnarray*}
where $x\in\mathbb{R}$ is the state of the controller,
\begin{align}
u=\begin{bmatrix}u_1 \\ u_2\end{bmatrix}, \ \
C=\begin{bmatrix}T_{\rm I}/T_{stp} \\ -T_{\rm I}/T_{stp}\end{bmatrix}, \ \
D=\begin{bmatrix}T_{\rm P} \\ -T_{\rm P}\end{bmatrix},\ \ 
\beta=\begin{bmatrix}5.5 \\ 5.5\end{bmatrix}. 
\nonumber
\end{align}
$T_{\rm P}$ is a proportional gain and $T_{\rm I}$ is an integral gain.
$T_{\rm P}$ and $T_{\rm I}$ were respectively set to 5.45 and 1.55 for angle tracking control and to 7.45 and 4.75 for torque tracking control; these values were determined by trial-and-error. 
$\beta$ is a bias determined by the PDCV specifications.
Tracking control experiments with respect to the joint angle and torque were conducted under step-like references set within a range of $\pm20$$^\circ$ and $\pm2.0$ Nm, respectively.
Fig.\,\ref{fig:sensorless_control} shows the resulting responses. Further, the actual tracking errors $\zeta_\psi:=\bar{\psi}-\psi$ and $\zeta_\tau:=\bar{\tau}-\tau$ are evaluated using the measured signals $\psi$ and $\tau$ for the validation.
It should be noted that the feedback error $e$ differs from the actual tracking errors $\zeta_\psi$ and $\zeta_\tau$.

\subsubsection{Angle-sensor-less control}
Fig.\,\ref{fig:sensorless_control}\subref{subfig:angle_control} shows the result of encoder-less control.
The upper graph shows the time responses with regard to the angle, where the black dashed line indicates the reference; the red line, the estimation; and the green line, the sensor value.
The bottom graph shows the tracking error $\zeta_\psi$.
The steady-state error is seen to be less than 2.1$^\circ$ at around 60 s, where the worst-case steady-state error is $2.1/40= 0.0525$ $(5.25 \%)$.
Therefore, the UKF-based encoder-less angle control system is confirmed to achieve steady-state tracking control performance of $94.75 \%$.

\subsubsection{Torque-sensor-less control}
\begin{figure}[t]
		\centering
		\subfloat[Time responses of measured and estimated joint angles (upper) and tracking error $\zeta_\psi$ between the reference and the measured angle (lower).]
		{\includegraphics[width=1.0\hsize]{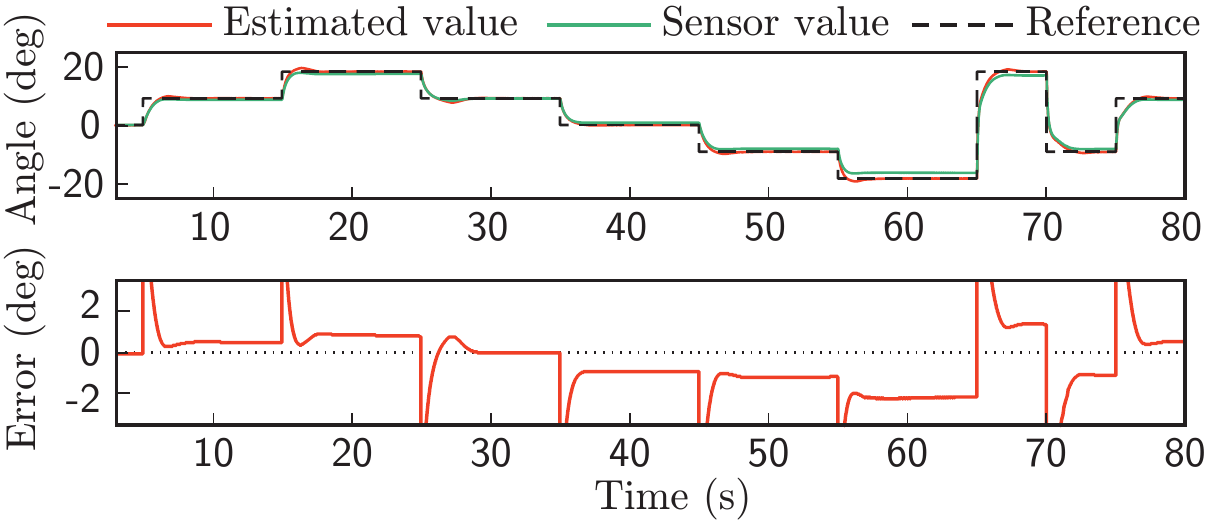}
		\label{subfig:angle_control}}\\
		\subfloat[Time responses of measured and estimated torque (upper) and tracking error $\zeta_\tau$ between the reference and the measured torque (lower).]
		{\includegraphics[width=1.0\hsize]{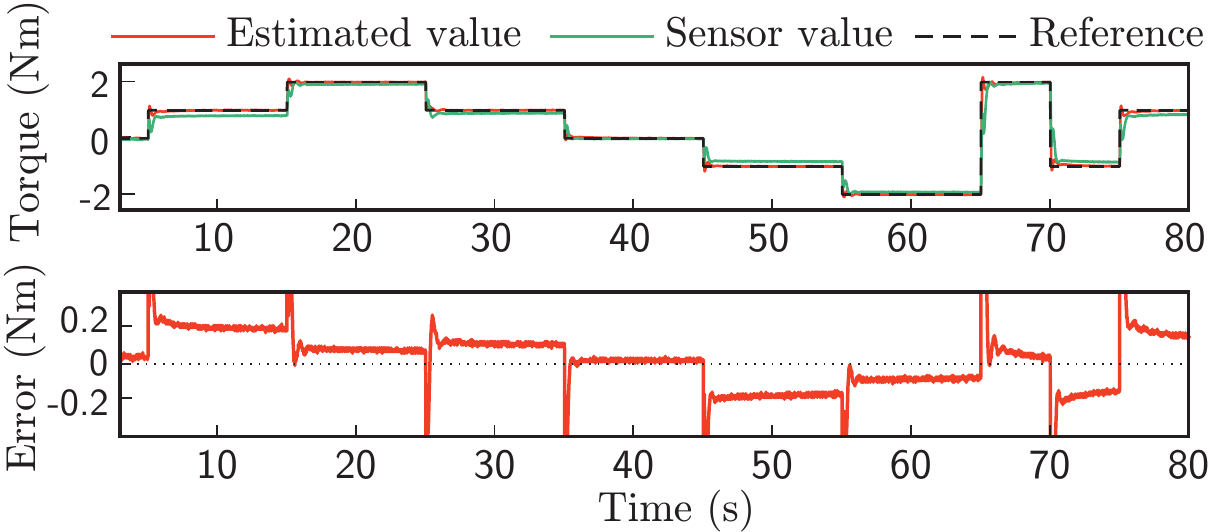}
		\label{subfig:torque_control}}
		\caption{Experimental results of UKF-based sensor-less tracking control.}
		\label{fig:sensorless_control}
	\end{figure}

Fig.\,\ref{fig:sensorless_control}\subref{subfig:torque_control} shows the result of torque-sensor-less control.
The upper graph shows the time responses with regard to the torque; the line colors have the same meaning as those in Fig.\,\ref{fig:sensorless_control}\subref{subfig:angle_control}.
The bottom graph shows the tracking error $\zeta_\tau$.
The steady-state error is seen to be less than $0.2$ Nm at around 10 and 50 s, where the worst-case steady-state error is $0.2/4.0=0.05$ $(5.0 \%)$.
Therefore, the UKF-based sensor-less torque control system is confirmed to achieve steady-state tracking control performance of $95.0 \%$.

\section{Conclusion}\label{sec:Conclusion}
This paper proposed a detailed nonlinear mathematical model of an antagonistic PAM actuator system that can estimate the joint angle and torque with a UKF.
The proposed model is described in a hybrid state-space representation. It consists of the contraction force of the PAM, joint dynamics, fluid dynamics of compressed air, mass flow of the valve, and friction models.
The friction models include the novel form of the Coulomb friction that depends on the inner pressure of the PAM.
For model validation, offline and online estimation using the experimental data and sensor-less tracking control with regard to the joint angle and torque of the rotational joint were conducted to evaluate the estimation accuracy and tracking control performance.
The following results were obtained:
(1) In the offline simulation, the UKF-based estimation achieved an estimation accuracy of 6.13\% and 4.94\% for the joint angle and torque, respectively.
(2) The computer on which the UKF was implemented could estimate the joint angle and torque in real time, for which the UKF achieved worst-case estimation errors of 7.91 \% and 6.01 \%, respectively.
(3) From sensor-less control applications, the UKF-based sensor-less control systems were confirmed to achieve steady-state tracking control performance of more than 94.75 \% and 95.00~\% with respect to the joint angle and torque, respectively.

Future studies will address problems such as the development of an antagonistic PAM actuator system for enabling sensor-less stiffness control, constrained control for a lightweight and flexible actuator, and development of a safe and force-interactive PAM actuator system for estimating the reaction torque against humans or the environment.

\appendices

\section*{Acknowledgment}
This work was supported by JSPS KAKENHI Grant Numbers JP25709014 and JP18K04012.
The authors thank Ms. Rieko Kadoya for data collection and the helpful discussions.

\ifCLASSOPTIONcaptionsoff
  \newpage
\fi

\bibliographystyle{IEEEtran}
\bibliography{reference}

\begin{IEEEbiography}[{\includegraphics[width=1in,height=1.25in,clip,keepaspectratio]{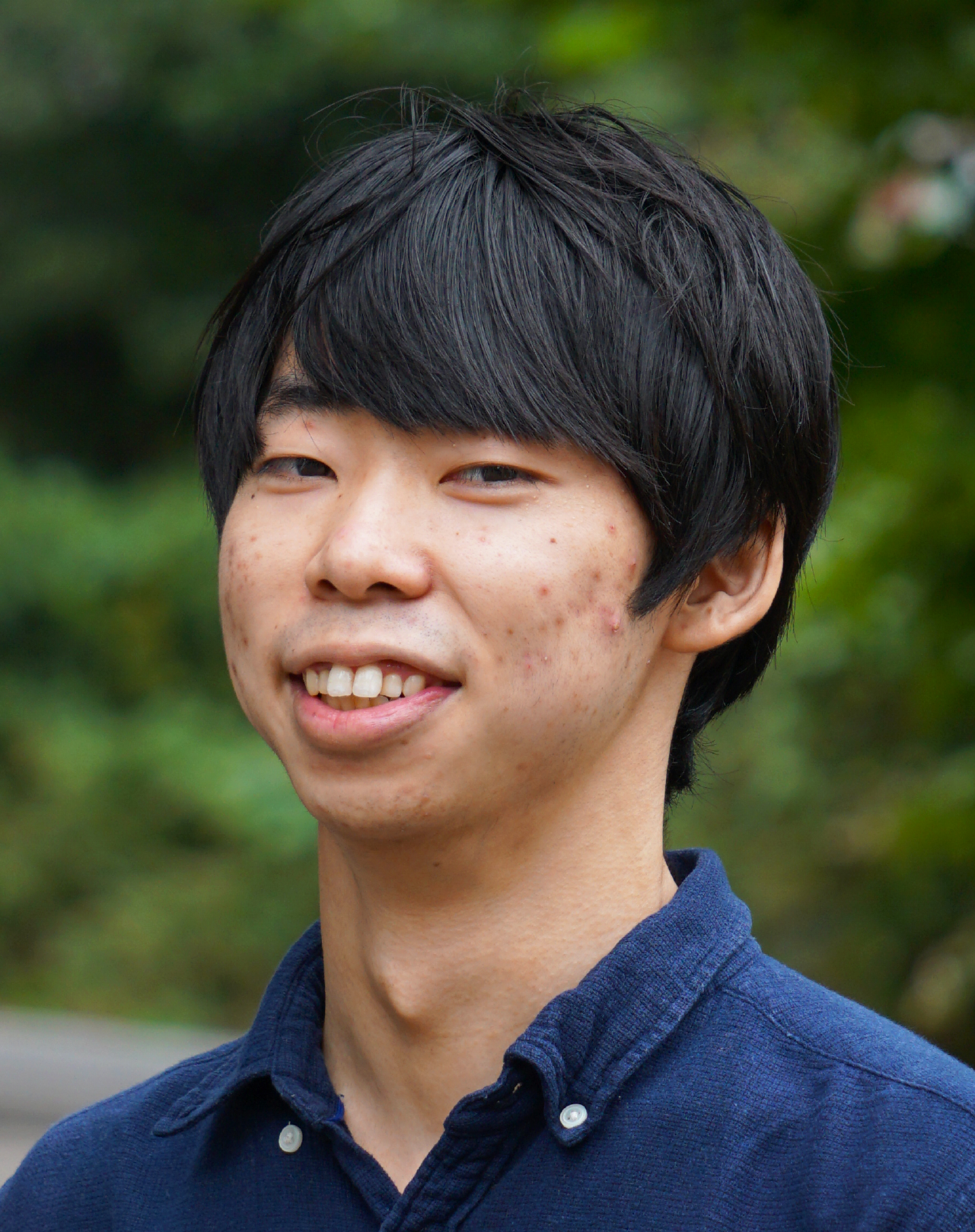}}]{Takaya Shin}
received the B.S. degree in Informatics and Engineering from The University of Electro-Communications, Tokyo, Japan, in 2020. He is currently an M.S. student at The University of Electro-Communications, Tokyo, Japan.

His research interests include control applications and modeling/control of pneumatic artificial muscles.
\end{IEEEbiography}

\begin{IEEEbiography}[{\includegraphics[width=1in,height=1.25in,clip,keepaspectratio]{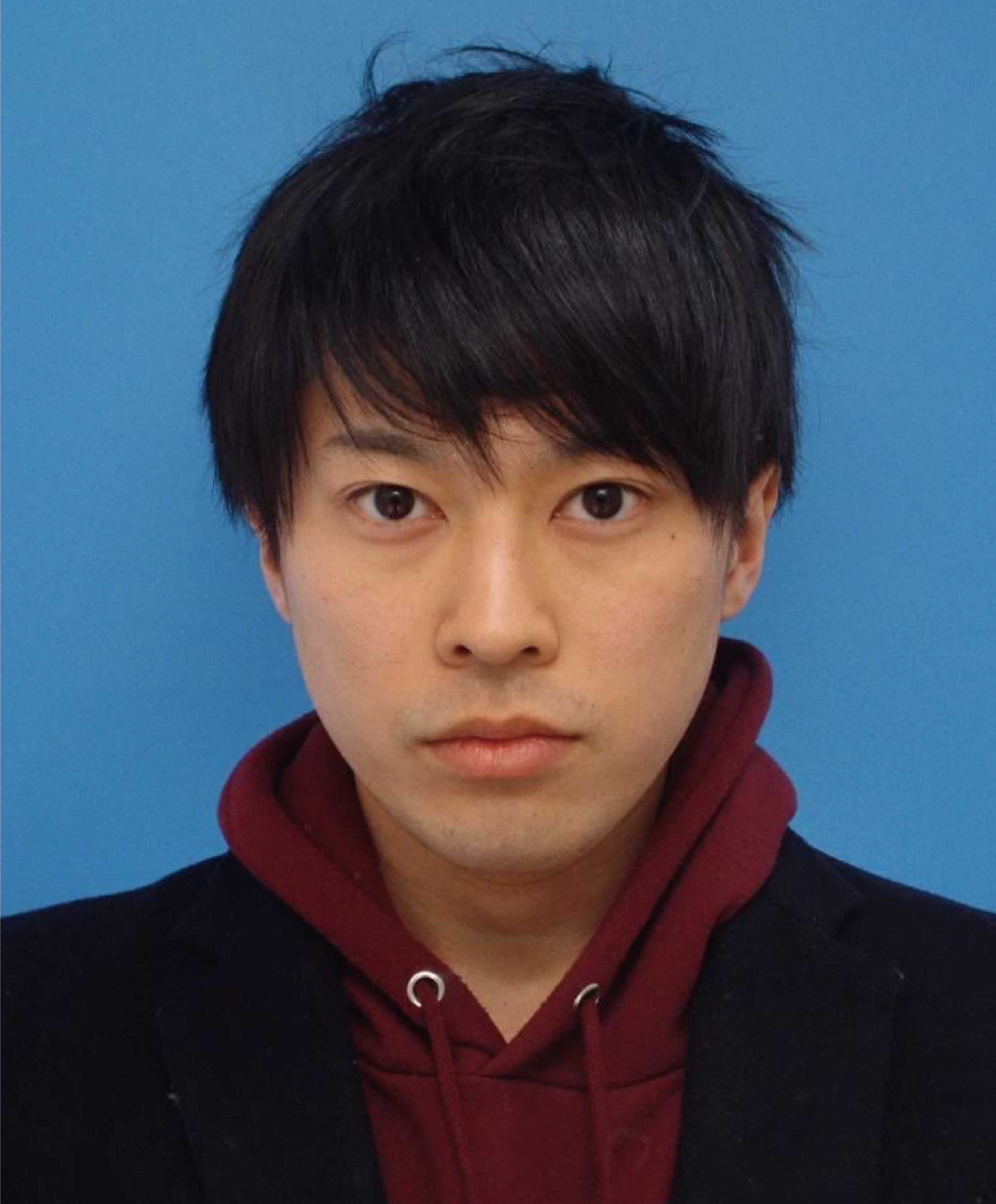}}]{Takumi Ibayashi}
received the B.S. and M.S. degrees in Mechanical Engineering from The University of Electro-Communications, Japan, in 2018 and 2020, respectively.

He joined Safie, Inc., in 2020. 
His research interests include modeling and control of pneumatic artificial muscles.
\end{IEEEbiography}

\begin{IEEEbiography}[{\includegraphics[width=1in,height=1.25in,clip,keepaspectratio]{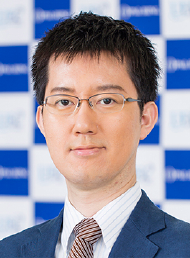}}]{Kiminao Kogiso}
received the B.S., M.S., and Ph.D. degrees in Mechanical Engineering from Osaka University, Japan, in 1999, 2001, and 2004, respectively.

He was a postdoctoral researcher of the 21st Century COE Program and became an Assistant Professor in Department of Information Systems, Nara Institute of Science and Technology, Nara, Japan, in 2004 and 2005, respectively.
Since March 2014, he has been an Associate Professor in Department of Mechanical Engineering and Intelligent Systems, The University of Electro-Communications, Tokyo, Japan.
From November 2010 to December 2011, he was a visiting scholar at the Georgia Institute of Technology, GA, USA.

His research interests include constrained control, control of decision makers, cyber-security of control systems, and their applications. 
\end{IEEEbiography}

\end{document}